\documentclass[useAMS,usenatbib]{mn2e}

\pdfoutput=1
\usepackage{times}  
\usepackage{listings}
\usepackage{graphics}
\usepackage{subfigure}
\usepackage{graphicx}
\usepackage{amssymb}
\usepackage{hyperref}
\usepackage{aas_macros}
\usepackage{verbatim}

\long\def\symbolfootnote[#1]#2{\begingroup%
\def\thefootnote{\fnsymbol{footnote}}\footnote[#1]{#2}\endgroup} 

\newcommand{\GALFORM}{{\tt GALFORM}}
\newcommand{\GADGET}{{\tt GADGET2}}
\newcommand{\SUBFIND}{{\tt SUBFIND}}
\newcommand{\Mpc}{{\rm Mpc}}
\newcommand{\Msolyr}{{\rm M_{\odot}yr^{-1}}}
\newcommand{\Msol}{{\rm M_{\odot}}}
\newcommand{\AB}{{\rm AB}}

\input{epsf} \title{Lightcone mock catalogues from semi-analytic
models of galaxy formation - I. Construction and application to the
BzK colour selection}

\author[Merson {\it et al.}]
       {\parbox[h]{\textwidth}{Alexander~I.~Merson$^{1,2}$\thanks{E-mail:
       a.i.merson@durham.ac.uk}, Carlton~M.~Baugh$^1$,
       John~C.~Helly$^1$, Violeta~Gonzalez-Perez$^1$, Shaun~Cole$^1$,
       Richard~Bielby$^1$, Peder~Norberg$^1$, Carlos~S.~Frenk$^1$,
       Andrew~J.~Benson$^3$, Richard~G.~Bower$^1$,
       Cedric~G.~Lacey$^1$, Claudia~del~P.~Lagos$^1$}
  \vspace*{3pt}\\
  \noindent $^1$Institute for Computational
  Cosmology, Department of Physics, University of Durham, South Road,
  Durham DH1 3LE, UK\\ $^2$Department of Physics and Astronomy, University College
  London, Gower Street, London WC1E 6BT\\ $^3$Carnegie Observatories, Pasadena, CA, U.S.A}

\date{}

\pagerange{\pageref{firstpage}--\pageref{lastpage}}
\pubyear{2012}

\begin{document}

\maketitle
\title{Lightcone mock catalogues}
\label{firstpage}

\begin{abstract}
We introduce a method for constructing end-to-end mock galaxy
catalogues using a semi-analytical model of galaxy formation, applied
to the halo merger trees extracted from a cosmological N-body
simulation. The mocks that we construct are \emph{lightcone}
catalogues, in which a galaxy is placed according to the epoch at
which it first enters the past lightcone of the observer, and
incorporate the evolution of galaxy properties with cosmic time. We
determine the position between the snapshot outputs at which a galaxy
enters the observer's lightcone by interpolation. As an application,
we consider the effectiveness of the BzK colour selection technique,
which was designed to isolate galaxies in the redshift interval
$1.4<z<2.5$. The mock catalogue is in reasonable agreement with the
observed number counts of all BzK galaxies, as well as with the
observed counts of the subsample of BzKs that are star-forming
galaxies. We predict that over $75$ per cent of the model galaxies
with ${\rm K_{\AB}}\leqslant 23$, and $1.4<z<2.5$, are selected by the
BzK technique. Interloper galaxies, outside the intended redshift
range, are predicted to dominate bright samples of BzK galaxies
(i.e. with ${\rm K_{\AB}}\leqslant 21$). Fainter ${\rm K}$-band cuts
are necessary to reduce the predicted interloper fraction. We also
show that shallow ${\rm B}$-band photometry can lead to confusion in
classifying BzK galaxies as being star-forming or passively
evolving. Overall, we conclude that the BzK colour selection technique
is capable of providing a sample of galaxies that is representative of
the $1.4<z<2.5$ galaxy population.
\end{abstract}


\section{Introduction}
\label{sec:intro}
Modern galaxy surveys such as the Sloan Digital Sky Survey
\citep[SDSS,][]{York00} and the 2-degree Field Galaxy Redshift Survey
\citep[2dFGRS,][]{Colless01,Colless03} have revolutionised our view of the 
galaxy distribution and have played a key role in shaping 
the constraints on our cosmological model \citep[e.g.][]{Norberg01,
Norberg02a,Zehavi05b,Cole05,Sanchez06,Tegmark06,Sanchez09,Zehavi11,
Sanchez12*}. 
The size of these surveys has heralded the start of an era of 
precision cosmology wherein we can measure statistics, such 
as the galaxy luminosity function, with random errors that 
are smaller than the systematic errors. To continue to make progress 
it is essential that we improve our understanding of how the 
estimation of such statistics is affected by the construction 
of a galaxy survey and the selection criteria applied. 
Mock galaxy catalogues, which mimic the selection effects in 
real surveys, have emerged as an essential tool with which 
to achieve this aim, and play a central role in the analysis 
and exploitation of galaxy surveys.

When working with an observational galaxy catalogue, an estimator designed
to recover a statistic, such as the luminosity function or correlation
function, will have to compensate for a variety of effects such as
non-uniform coverage of the sky and a selection function that varies
strongly with radial distance from the observer. The primary advantage
of a mock catalogue is that, by construction, we already know the
`true' answer for the statistic without these effects. By comparing a
measurement extracted from a synthetic mock catalogue with the ideal
result (i.e. the statistic measured using a complete sample of
galaxies from the original simulation cube), one can adjust and tune
the performance of the estimator to reduce any systematic effects. A
prime example is that of algorithms designed to find groups of
galaxies, the calibration of which requires foreknowledge of the
underlying dark matter halo distribution in order to test how
faithfully the algorithm can recover these structures when working in
redshift space \citep{Eke04a, Robotham11, Murphy12}. Additionally,
mock catalogues can be used to forecast the scientific return of
future galaxy surveys \citep{Cole98, Cai09, Orsi10}. Therefore they
can help shape the design of a survey by assessing the level, and
quality, of the statistics recoverable with any particular
configuration. Finally, mock catalogues allow us to cast the predictions of
theoretical models of galaxy formation in a form that can be directly
compared against observables.

In this paper we present a method for building mock catalogues for
galaxy surveys, which can cover any redshift range. Mock catalogues
constructed in this way have already been used extensively by the
Galaxy And Mass Assembly (GAMA) survey
\citetext{\citealp{Driver09}, see
also \citealp{Robotham11,Alpaslan12}}. Here, we illustrate the power
of mock catalogues by evaluating the performance of the BzK colour
selection technique, which was designed to isolate galaxies in the
redshift interval $1.4<z<2.5$ \citep{Daddi04b}. This redshift range is
an exciting one for galaxy assembly, since it is thought that most of
the stellar mass of many of the progenitors of present day massive
galaxies formed during this period \citep{Madau98,
Dickinson03}. Unfortunately this epoch lies within the `redshift
desert' where the spectroscopic measurement of galaxy redshifts is
difficult due to the lack of strong spectral features at optical
wavelengths. Only recently, with the development of near-infrared
(NIR) spectroscopy, have large galaxy surveys begun to probe this
region and to assess the build-up of galaxies over this crucial period
\citep[e.g.][]{Franx03, vanDokkum03}.

Prior to this, knowledge of the galaxy population in the `desert' was
derived from photometry. This led to the development of colour
selection techniques designed to efficiently identify targets for
spectroscopic follow-up (which is much more expensive). A well-known
example of this is the Lyman-break dropout technique, proposed by
\cite{Steidel96a,Steidel03,Steidel04}, which identifies star-forming
galaxies at $z\sim3-10$ according to their rest-frame ultraviolet
colours and sampling of the Lyman break spectral feature. Other
examples include extremely red objects at $z\sim1$
\citep{Elston88,PJMcCarthy04} and distant red galaxies at $z\sim2$
\citep{Franx03}.

A popular photometric technique, designed to simultaneously identify
populations of star-forming and passively evolving galaxies, is the
BzK colour-criterion \citep{Daddi04b}. This approach, which selects
galaxies based on their $(B-z)$ and $(z-K)$ colours, is designed to
deliver galaxy samples within the redshift range $1.4<z<2.5$ that are
not biased by the presence of dust or by the age of their stellar
populations
\citep{Kong06,Hayashi07,Hayashi09,Grazian07,Hartley08,Lin11*}.

Early studies of ${\rm K_{\AB}}\lesssim 22$\footnote{\cite{Daddi04b}
originally used a ${\rm K_{Vega}}\leqslant 20$ selected sample. We have
converted this to the AB system using the ${\rm K}$-band conversion
from \cite{Blanton07}, where $m_{\AB}-m_{{\rm Vega}}=1.85$.}
BzK-selected galaxies revealed them to typically have large stellar
masses, $\sim10^{11}h^{-1}\Msol$, and, in the case of those labelled
as star-forming, high star-formation rates, $\sim100h^{-1}\Msolyr$
\citep{Daddi04b,Daddi04a,Daddi05b,Daddi05a,Reddy05,Kong06}. Such
properties, combined with high metallicities
\citep[e.g. ][]{Daddi04b,Hayashi09} and indications that these systems
are strongly clustered \citep{Hayashi07,Hartley08,Blanc08}, have led
many authors to speculate that BzK galaxies are the high-redshift
precursors of massive early-type galaxies found in groups and clusters
at the present day. The key question we address here is: Are the
properties of the bright galaxies identified by this selection
technique representative of the overall population with $1.4<z<2.5$,
or are we really just seeing a special subclass of galaxies?  If the
latter is true, is this simply because current observations have not
been sufficiently deep to see the fainter, more representative
galaxies or is the BzK criterion somehow biased towards selecting a
subset of the galaxy population? A galaxy mock catalogue is a vital
resource in helping to answer these questions by allowing an
assessment of the effects of observational selection on the
completeness of the galaxy sample as well as an examination of whether
or not the BzK criterion is sensitive to the intrinsic properties of
galaxies.

The layout of the paper is as follows: In
section~\ref{sec:mock_construction_overview} we summarise the various
methods used to construct mock galaxy catalogues. In section
~\ref{sec:model}, we introduce the numerical simulation and galaxy
formation model that we will use as inputs, before, in
section~\ref{sec:lightcone_construction}, providing further details of
our method for constructing lightcone mock catalogues. This section
provides a full overview of the lightcone construction, including the
assignment of positions to model galaxies. In
section~\ref{sec:BzK_section} we use a lightcone mock catalogue to
assess the performance of the BzK selection technique. Note that this
application only uses some features of the lightcone; the clustering
of BzK galaxies will be dealt with in a separate paper.  Finally, in
section~\ref{sec:conclusions}, we summarise our method and present our
conclusions. Throughout this paper we use magnitudes in the AB system.

\section{Constructing mock catalogues}
\label{sec:mock_construction_overview}

In this section we provide an overview of the basic procedure for
constructing mock catalogues and set out the advantages of using a
semi-analytical model of galaxy formation.

\subsection{Overview of the technique}
\label{sec:overview}
A very basic mock catalogue could be constructed by randomly sampling
one of the measured statistical distributions that describe the galaxy
population (e.g. the luminosity function or stellar mass
function). Although the resulting catalogue of galaxies would match
that particular statistic (by construction), without any further
information about the galaxies, such as their colour or spatial
distribution, the mock would be very limited.

Building a more realistic mock catalogue, with positional information
and including other galaxy properties and their evolution, requires
the use of a numerical simulation which follows the growth of
structure in the dark matter. The procedure for constructing mock
catalogues from a numerical simulation can be broken down into the
following steps: (i) generate a population of galaxies either
empirically or using a physical model, using either the dark matter
distribution or dark matter halos, (ii) place these galaxies into a
cosmological volume, (iii) apply the angular and radial selection
functions of the survey.

\subsubsection{Generating a galaxy population}
\label{sec:generating_a_galaxy_pop}
To generate a population of galaxies one must first model the
distribution of dark matter, which is often done with a N-body
simulation. Dark matter only N-body simulations allow us to build halo
populations using gravity alone. The full spatial information provided
by N-body simulations allows one to extract clustering information,
which would otherwise not be available if a Monte-Carlo approach was
to be used. Additionally, the merger histories of halos in N-body
simulations will also include environmental effects, such as halo
assembly bias \citep{Gao05}.

The way in which dark matter halos are populated with galaxies is
where the methods of mock catalogue construction can
differ. \cite{Blaizot05} \citep[see also ][]{Baugh08} summarise
several of the different methods available, which include using
phenomenological models to assign galaxies to dark matter particles in
the simulation \citep[e.g. ][]{Cole98} or using empirically derived
statistics, such as the \emph{halo occupation distribution}
\citep[HOD, ][]{BW02, Song12} or \emph{sub-halo abundance matching}
\citep{VO04}. Other, more physical approaches are also possible. For
instance, one could include the baryons in the original simulation,
using either a grid-based or particle-based method to solve the
hydrodynamical equations. The problem with direct, hydrodynamical
simulations however, is that they are computationally expensive and
so, in practice, are restricted to small volumes \citep[e.g. the $25
h^{-1}\Mpc$ and $100 h^{-1}\Mpc$ boxes used in the Overwhelmingly Large
Simulations project of ][]{Schaye10}.

A powerful approach, that we choose to adopt, is to use a
\emph{semi-analytical model} of galaxy formation to populate the halo
merger trees extracted from a high resolution, cosmological N-body
simulation
\citep{Kauffmann99c,Benson00b,Blaizot05,KW07,Sousbie08,Overzier09,Cai09,Henriques12,Overzier12*}.
Modelling of various physical processes, such as the cooling of gas
within dark matter halos, is necessary to follow the baryonic
component and predict the fundamental properties of galaxies, such as
their stellar mass and star formation history. The adoption of an
\emph{initial mass function} (IMF), a \emph{stellar population
synthesis} (SPS) model and a treatment of dust extinction allows these
fundamental properties to be connected with observables, thus enabling
a direct comparison between observations and the predictions of the
galaxy formation model.

\subsubsection{Generating a cosmological volume}
\label{sec:simulation_size}

Current and future galaxy surveys are designed to probe ever larger
cosmological volumes. As a result there is a growing demand for
simulations with boxes of sufficient size to match the volumes of
these surveys. Unfortunately, current computing power means that a
compromise must often be made between the volume of the simulation box
and the resolution at which the simulation is carried out. Therefore a
sufficiently large cosmological volume can only be sampled by tiled
replication of a smaller box simulation. For very shallow galaxy
surveys (e.g. with a median redshift $z \lesssim 0.05$), the lookback
time is sufficiently small that typical galaxy properties will not
have undergone significant evolution across the redshift interval
covered by the survey. In these instances, the statistics of the
galaxy population at the extremes of the survey will not be too
dissimilar to the statistics today and so one can build a mock
catalogue using galaxies from a single simulation snapshot. However,
for very deep galaxy surveys which cover a significant lookback time,
we would expect to see substantial evolution in galaxy properties and
in the growth of large-scale structure. Therefore more sophisticated
mock catalogues, that tile the survey volume using many different
simulation snapshots, are required to adequately reproduce the
evolution seen in the properties of galaxies and their clustering. The
mock catalogues that we construct in this work are \emph{lightcone}
mock catalogues, in which galaxies are placed according to the epoch
at which they first cross the observer's past lightcone, i.e. at the
location at which the light emitted from the galaxy has just enough
time to reach the observer, and thus incorporate the evolution of
structure with cosmic time.

Finally, observations will be subject to uncertainties or biases,
introduced as a result of survey design or selection effects, and so
to properly relate theoretical predictions to observations we must
subject the simulated data to the same selection functions as the
observed galaxy sample.

\subsection{Why use a semi-analytical galaxy formation model?}
\label{sec:sams_reasons}
Modelling the formation of galaxies is a difficult task. Part of the
problem is that our knowledge of the underlying physics is limited and
so we cannot simply write down a precise formulation for every
process. Furthermore, despite the continued development of direct,
hydrodynamic simulations, current computational capabilities mean that
many of the relevant processes (for example star formation or
supernova feedback) remain firmly below the resolution limits of
direct simulations and can only be addressed through ``sub-grid''
physics. Semi-analytic models describe the sub-grid physics using
physically motivated, parametrised equations that follow the
evolution of baryons trapped in the gravitational potential wells of
hierarchically grown dark matter halos
(\citealt{WR78,Cole91,WF91,Kauffmann93,Cole94}; for reviews of the
semi-analytical approach see \citealt{Baugh06} and
\citealt{Benson10a}). There are several compelling advantages to using
semi-analytic models for building mock catalogues:

\begin{enumerate}
\item The development of deep, wide-field photometric galaxy surveys
spanning large cosmological volumes has led to demand for large
(suites of) mock catalogues that can be constructed rapidly and
accurately. Semi-analytic modelling is currently the only physical
approach that meets these ideals: such models are capable of
populating large cosmological volumes with galaxies much faster and at
a lower computational cost than is currently possible with
hydrodynamical simulations.

\item The modular design of semi-analytic models allows new physics to
be incorporated readily. Combined with their short run-time, this
means that semi-analytic models can be tuned to match observations
quickly, in response to a change to the background cosmology or to the
galaxy formation physics. Moreover, the larger computational box that
can be used in the N-body and semi-analytical approach, compared with
a hydrodynamical simulation, means that the clustering predictions are
robust out to larger scales.

\item Empirically motivated methods, such as HOD modelling, must first
be calibrated against observational data, \citep[e.g. the LasDamas
mock catalogues, ][]{McBride09}. Hence, mock catalogues built using
such methods are limited by the availability of observational data at
high redshift. Furthermore, the data that is available may be affected
by sample variance leading to unrepresentative HOD parameters being
fitted. Semi-analytic models however, once tuned to fit the
observations of galaxies at low redshift, can {\it predict} galaxy
properties out to high redshift, without further observational
input. Mock catalogues built from semi-analytical models are therefore
much more flexible than catalogues constructed using other methods.

\item The next generation of galaxy surveys will map the sky across a
large portion of the electromagnetic spectrum, with multi-wavelength
follow-up observations resulting in potentially complex survey
selection functions, such as for the Galaxy And Mass Assembly (GAMA)
Survey \citep{Driver11}. Ideally mock catalogues for future surveys
need to provide a diverse range of galaxy properties as well as
providing the capability to select galaxies simultaneously in multiple
bands. Semi-analytic models model the complete star formation history
for each galaxy and so can predict many different galaxy
properties. Mock catalogues based upon semi-analytic models are
already capable of mimicking sophisticated multi-band selection
criteria.

\end{enumerate}

\section{Galaxy Formation Model} 
\label{sec:model}
The model we adopt to generate the galaxy population for our mock
catalogues is the \cite{Bower06} variant of the \GALFORM{}
semi-analytic model \citep{Cole00}. To build realistic lightcone mock
catalogues we require spatial information, so we use dark matter halo
merger trees extracted from the \emph{Millennium Simulation}
\citep{Springel05a}.

\subsection{The Millennium Simulation}
\label{sec:Millennium}

\subsubsection{Cosmology and parameters}
\label{sec:Millennium_cosmology}
The population of dark matter halos for our mock catalogues is
provided by the \emph{Millennium Simulation}, a $2160^3$ particle
N-body simulation of the ${\rm \Lambda}$CDM cosmology carried out by
the Virgo Consortium \citep{Springel05a}. This simulation follows the
hierarchical growth of cold dark matter structures from redshift
$z=127$ through to the present day in a cubic volume of size
$500h^{-1}\Mpc$ on a side. Halo merger trees are constructed using
particle and halo data stored at 64 fixed epoch snapshots that are
spaced approximately logarithmically in expansion factor. The
Millennium trees have a temporal resolution of approximately $0.26\,
{\rm Gyr}$ at the present day, with approximate resolutions of
$0.38,0.35,0.26\,{\rm Gyr}$ at redshifts $z=0.5,1,2$
respectively. Halos in the simulation are resolved with a minimum of
20 particles, corresponding to a halo resolution of $M_{{\rm
halo,lim}} = 1.72\times10^{10}h^{-1}\Msol$, significantly smaller than
expected for the Milky Way's dark matter halo.

The cosmological parameters adopted in the \emph{Millennium
Simulation} are: a baryon matter density $\Omega_{{\rm b}} = 0.045$, a
total matter density $\Omega_{{\rm m}} = \Omega_{{\rm b}} +
\Omega_{{\rm CDM}} = 0.25$, a dark energy density $\Omega_{{\rm
\Lambda}} = 0.75$, a Hubble constant $H_0 = 100h \,{\rm km\,
s}^{-1}\Mpc^{-1}$ where $h = 0.73$, a primordial scalar spectral index
$n_{\rm s}=1$ and a fluctuation amplitude $\sigma_{8}=0.9$. These
parameters were chosen to match the cosmological parameters estimated
from the first year results from the Wilkinson Microwave Anisotropy
Probe \citep{Spergel03}.

\subsubsection{Construction of halo merger trees}
\label{sec:Millennium_trees}

To construct the halo merger trees one must first identify groups of
dark matter particles in each of the simulation snapshots. This is
done using the \emph{Friends-Of-Friends} algorithm \citep[FOF,
][]{Davis85}. The \emph{Millennium Simulation} was carried out with a
specially modified version of the \GADGET{} code
\citep{Springel05b} with a built-in FOF group-finder, allowing FOF
groups to be identified on the fly. The algorithm \SUBFIND{}
\citep{Springel01} was then used to identify self-bound, locally
over-dense sub-groups within the FOF groups. This procedure typically
results in the bulk of the mass of a FOF group being assigned to one
large sub-group which represents the background mass distribution of
the halo. The remaining mass is usually split between smaller
satellite sub-groups orbiting within the halo and unbound ``fuzz''
particles which are not associated with any sub-group.

However, it is not uncommon for the FOF algorithm to join together
structures which might be better considered to be separate halos for
the purposes of semi-analytic galaxy formation. For example, nearby
groups may be linked by tenuous ``bridges'' of particles or they may
only temporarily be joined. The merger tree algorithm we use in this
work is intended to deal with these cases and ensure that the
resulting trees are strictly hierarchical, i.e. once two halos are
deemed to have merged they should remain merged at all later times.

The first step in the construction of the merger trees is to identify
a descendant for each sub-group at the next snapshot. The descendant
of each sub-group is identified as the sub-group at the next snapshot
that contains the largest number of the $N_{{\rm link}}$ most bound
particles, where
\begin{equation}
N_{{\rm link}} = {\rm max}\left ( f_{{\rm trace}} N_{{\rm
p}},N_{{\rm linkmin}}\right ),
\label{eq:Nbound}
\end{equation}
with $N_{{\rm p}} \geqslant 20$, as already stated, and $f_{{\rm
trace}}$ and $N_{{\rm linkmin}}$ are set to 0.1 and 10
respectively. Defining $N_{{\rm link}}$ in this way means that in well
resolved cases we follow the most bound ``core'' of the sub-group,
which is important for satellite sub-groups which may be tidally
stripped of their outer parts. For the smallest groups with $N_{{\rm
p}} \sim 20$, $N_{{\rm link}}=10$ so we are following up to $50\%$ of
the particles and so preventing inaccurate assignment due to low
number statistics.

The \SUBFIND{} algorithm occasionally temporarily ``loses'' a
sub-group between snapshots. For example, a sub-group may be
identified at snapshot $i$, lost at one or more subsequent snapshots,
and then identified again at snapshot $i+n$, where $n > 1$. This can
happen if a small, isolated group briefly falls below the resolution
limit or if a satellite sub-group passes close to the centre of its
host halo. In either case we would like to identify the sub-group at
snapshot $i+n$ as the descendant of the sub-group at snapshot $i$. Our
approach to achieve this aim is as follows:
\begin{enumerate}
\item Identify sub-groups which may have been lost by \SUBFIND{}.
\item Identify sub-groups which may have just been reacquired by
\SUBFIND{}.
\item Attempt to locate descendants of the sub-groups in (i) with the
sub-groups in (ii).
\end{enumerate}

Groups which are ``lost'' are identified by looking for groups which
either have no immediate descendant or are not the most massive
progenitor of their immediate descendant. Some of these groups will
have been lost because they have genuinely been disrupted and absorbed
into the parent halo, but some will reappear later. Groups which have
just been reacquired are identified by looking for ``orphan''
sub-groups, i.e. groups with no immediate progenitors.

For each lost sub-group at snapshot $i$ (where $i$ is not the present
day), we examine the orphan sub-groups at snapshot $i+2$, $i+3$, ... ,
$i+N_{step}$. An orphan sub-group is identified as the descendant of a
lost sub-group if at least a fraction $f_{{\rm link}}$ of the $N_{{\rm
link}}$ most bound particles from the lost group are in the orphan
group and no orphan descendant can be found at earlier snapshots. We
usually set $N_{{\rm step}}=5$ and $f_{{\rm link}}=0.5$.

If this procedure results in the identification of a descendant for a
sub-group, then that descendant will be used in the subsequent stages
of the construction of the merger trees. For all other sub-groups the
descendant is taken to be the immediate descendant at the next
snapshot. For the construction of merger trees, having a sub-group and
its descendant separated by multiple snapshot outputs is not a
problem. However, this is inconvenient for codes, such as \GALFORM{},
which expect the descendant of a subhalo to always be found in the
next snapshot. To avoid this, for those subhalos that are temporarily
lost, interpolated sub-halos are inserted at each snapshot where the
sub-halo is `missing'. For very high resolution simulations this is a
common occurrence. However, for simulations like the \emph{Millennium
Simulation} such interpolated sub-halos are rare.

Next, the sub-groups at each snapshot are organised into a hierarchy
of halos, sub-halos, sub-sub-halos etc.. For each sub-group in a FOF
group we identify the least massive of any more massive ``enclosing''
sub-groups in the same FOF group. Sub-group A is said to enclose
sub-group B if B's centre lies within twice the half mass radius of
A. Any sub-group which is not enclosed by another is considered to be
an independent halo. We also consider a sub-group to be an independent
halo if it has retained at least 75 per cent of the maximum mass it
has ever had while being the most massive sub-group in its FOF
group. This is because we expect a halo involved in a genuine merger
with a more massive halo to be stripped of mass. In either case, if a
sub-group is deemed to be an independent halo then any sub-groups it
encloses are also assigned to that halo.

At this stage we have, for each snapshot, a population of halos, each
of which consists of a grouping of \SUBFIND{} sub-groups with
pointers linking each sub-group with its descendant at the next
snapshot. We choose the descendant of a halo to be the halo at the next
snapshot which contains the descendant of the most massive sub-group
in the halo. This defines the halo merger tree structure.

The \GALFORM{} model assumes that when a halo merges with another,
more massive ``host'' halo, that its hot gas is stripped away so that
no further gas can cool in the less massive halo. Since a halo can
only be stripped of hot gas once, we wish to treat these objects as
satellite sub-halos within their host halo for as long as they survive
in the simulation, even if their orbit puts them outside the virial
radius of their host halo at some later times. We therefore attempt to
identify cases where halos fragment, and re-merge them.

In practice we implement this by looking for satellite sub-groups
which split off their host to become independent halos at the next
snapshot. A sub-group will be re-merged if it satisfies all of the
following conditions:
\begin{itemize}
\item The sub-group is the most massive progenitor of its
  descendant. This is taken to mean that the sub-group survives at the
  next snapshot.
\item The sub-group is not the most massive sub-group in its
  halo. This indicates that it is a satellite sub-halo within a
  larger halo.
\item The descendant of the sub-group is the most massive group in its
halo.
\item The descendant of the sub-group belongs to a halo other than the
  descendant of the halo containing the original sub-group. This
  indicates that the host halo has fragmented.
\end{itemize}
The last condition is necessary because a sub-group can sometimes
become the most massive in its parent halo without any halo
fragmentation occurring, especially if the halo consists of two
sub-groups of similar mass. If these conditions are met, the halo
containing the descendant of the satellite sub-group is merged with the
descendant of the host halo.

Following this post-processing, we are left with, for the
\emph{Millennium Simulation}, approximately 20 million halo merger
trees with, in total, approximately 1 billion nodes.

\subsection{The GALFORM semi-analytic model}
\label{sec:GALFORM}
The Durham semi-analytical galaxy formation model, \GALFORM{},
originally developed by \cite{Cole00}, models the star formation and
merger history of a galaxy and makes predictions for many galaxy
properties including luminosities over a substantial wavelength range
extending from the far-UV through to the sub-millimetre
\citep{Baugh05,Lacey08,Lacey11,Lagos11b,Fanidakis11,Lagos12*}.

\subsubsection{Model overview}
\label{sec:galform_overview}

The \GALFORM{} model populates a distribution of dark matter halos
with galaxies by using a set of coupled differential equations to
determine how, over a given time-step, the ``subgrid'' physics regulate
the size of the various baryonic components of galaxies. \GALFORM{}
models the main physical processes governing the formation and
evolution of galaxies: (i) the collapse and merging of dark matter
(DM) halos, (ii) the shock-heating and radiative cooling of gas inside
DM halos, leading to the formation of galactic disks (iii) quiescent
star formation in galactic disks, (iv) feedback as a result of
supernovae, active galactic nuclei and photo-ionisation of the
inter-galactic medium, (v) chemical enrichment of stars and gas and
(vi) dynamical friction driven mergers of galaxies within DM halos,
capable of forming spheroids and triggering starburst events. The
prescriptions describing these physical processes are described in a
series of papers:
\cite{Cole00,Benson03,Baugh05,Bower06,Font08,Lacey08,Lagos11a}, as
well as in the reviews by \cite{Baugh06} and \cite{BB10}.

The star-formation history of a galaxy can be determined by tracking
the star-formation rate, SFR, and chemical enrichment predicted in its
progenitors. Convolving this with a model SSP (single stellar
population) allows one to predict the spectral energy distribution
(SED) of the galaxy, which in turn can be sampled by filter
transmission curves to predict rest-frame magnitudes in various
ultraviolet (UV), optical, near-infrared and far-infrared bands. Given
the redshift of a galaxy, the fixed filter transmission curves can be
shifted by an appropriate amount to obtain observer-frame magnitudes.
Dust extinction is incorporated by assuming the dust (whose mass is
predicted by the chemical evolution model) is mixed together with the
stars in the disk of the galaxy in two phases: in clouds and in a
diffuse component (see \citealt{Granato00}). Assuming a distribution
of dust grain sizes, and combining this with the predicted
scalelengths of the disk and bulge, allows one to calculate the
optical depth and apply the appropriate attenuation to the luminosity
at various wavelengths. We use the stellar population synthesis model
of \citeauthor{BC03} \citep[a version from 1999, that is described
  in][]{BC03} and assume a \cite{Kennicutt83} IMF in all modes of star
formation.

We set the adjustable parameters in \GALFORM{} by requiring that the
model predictions match a subset of observations, primarily of the
local galaxy population. We have traditionally assigned more weight in
this process to matching the optical and near-infrared luminosity
functions \citep[see e.g. ][ for a discussion of an automated version
of this process]{Bower10}. The requirement of matching the observed
luminosity function has led to the inclusion of different feedback
mechanisms to regulate star formation. Feedback from stellar winds and
supernovae (SNe) is important for reheating cold gas (and thus
quenching the star formation) in small halos. This process has been
shown to allow the models to reproduce the faint end of the observed
galaxy luminosity function \citep{Benson03}. The major extension made
in \cite{Bower06} was the introduction of feedback due to
active-galactic nuclei (AGN), which quenches the cooling flow in
massive, quasi-static hot halos and consequently shuts down star
formation in their central galaxies. This proposed solution to the
\emph{over-cooling problem} that had long plagued models of galaxy
formation \citep[e.g.][]{Benson03}, proved necessary to explain the
break in the luminosity function, allowing the model to reproduce the
${\rm b_J}$-band and K-band luminosity functions (including the
evolution at the bright end) out to redshift $z\sim 2$. In addition,
the \citeauthor{Bower06} model is able to accurately predict the
evolution of the galaxy stellar mass function out to $z\sim5$
\citep{Bielby11*}, successfully reproduce the clustering and abundance
of luminous red galaxies as seen in the SDSS \citep{Almeida08},
produce a bimodal distribution of galaxy colours that is in good
agreement with observations of the SDSS \citep{Gonzalez09} and match
the number counts and redshift distribution of extremely red objects
\citep{Gonzalez-Perez09}.

The \GALFORM{} calculation uses dark matter merger histories extracted
from the Millennium simulation. As commented upon above, this
simulation has 64 snapshots. Information about the baryonic content of
galaxies is tracked on finer timesteps, with 8 ``sub-steps'' inserted
between the N-body output times. Parts of the calculation, for example
the luminosity of the composite stellar population can be followed on
an even finer time grid, determined by an adaptive differential
equation solver.

\subsubsection{Placement of galaxies in halos}
\label{sec:placement_of_galaxies}

In the \GALFORM{} model, the treatment of the properties of a galaxy
will depend upon its status within its host halo, i.e. whether it is a
central or a satellite galaxy. Central galaxies are placed at the
centre of the most massive sub-halo of the host halo and are the focus
for all gas that is undergoing cooling. In the event of a halo merger,
we choose the central galaxy of the main (most massive) progenitor
halo as the central galaxy of the descendant halo, with any other
galaxies becoming satellites. It should be noted that according to
this definition, the central galaxy of each (sub)halo need not be the
most luminous or the one with the largest stellar mass.

Following a halo merger, the central galaxy of the less massive
progenitor halo becomes a satellite galaxy of the descendant halo. If
the most massive sub-halo of the less massive progenitor can no longer
be resolved (i.e. the sub-halo now has fewer than 20 particles and has
been lost), then the galaxy is placed on what was the most bound
particle in that sub-halo. Satellite galaxies are stripped of their
hot gas, thus quenching any further cooling and inhibiting long-term
star-formation \citep[see ][ for an alternative cooling model for
satellites]{Font08}. 

A merger timescale is calculated based upon the initial energy and
angular momentum of the satellite's orbit (which is chosen at random),
as well as the mass of the satellite and the mass of the halo hosting
the central galaxy and satellite system. It is expected that after
this time the effects of dynamical friction will have caused the
satellite to merge with the central galaxy. However, the merger
timescale of a satellite is recalculated every time the satellite's
host halo merges and becomes a sub-halo of a more massive halo
\citep[see ][]{Cole00}.

\section{Lightcone construction} 
\label{sec:lightcone_construction}
We now outline the method adopted to construct lightcone mock
catalogues. Our scheme shares many features in common with the methods
used by \cite{Blaizot05} and \cite{KW07}, with some improvements.

By first running the \GALFORM{} model\footnote{We stress that our
lightcone construction algorithms are independent of choice of
semi-analytic model and can be run using any input galaxy formation
model.} on the halo merger trees of the \emph{Millennium Simulation}
we generate a galaxy population that is used to build the lightcone
catalogues. Galaxy properties are stored for each fixed, snapshot
epoch that falls within the redshift range of interest for a
particular survey.

An observer is then placed inside the simulation box at a position
that can be set manually\footnote{Often one will choose to position
the observer manually if they desire the observer to be placed in a
specific location, such as an environment similar to the Local
Group.} or at random.

\subsection{Replication of the simulation box}
\label{sec:box_replication}

The cosmology used in the Millennium Simulation means that the
simulation box side-length, $L_{{\rm box}}=500h^{-1}\Mpc$, corresponds
to the co-moving distance out to $z=0.17$. Therefore, in order to
generate a cosmological volume that is of sufficient size to fully
contain any galaxy survey that extends out to a modest redshift, it is
necessary to tile replications of the simulation box (see the
discussion in $\S$~\ref{sec:simulation_size}).

The number of replications per axis, $n_{{\rm rep}}$, that need to be
stacked around the original box (containing the observer) is given
by,\footnote{$\lfloor x \rceil$ means that $x$ is rounded down to the
nearest integer.}
\begin{equation}
n_{{\rm rep}} = \left \lfloor \frac{r_{{\rm max}}}{L_{{\rm box}}}
\right \rceil + 1,
\label{eq:nrep}
\end{equation}
where $r_{{\rm max}}$ is the maximum co-moving radial distance that we
want to reach in the final mock catalogue. Including the original
simulation box, we have a total of $(2n_{{\rm rep}}+1)^3$
replications. The Cartesian co-ordinate system,
$(\mathbf{\hat{X}},\mathbf{\hat{Y}},\mathbf{\hat{Z}})$, of the
combined `super-cube' is then translated so that the observer is
located at the origin.

An unfortunate consequence of generating a large volume in this way is
that structures can appear repeated within the final lightcone
volume. Although repeated structures cannot have a co-moving separation
less than the simulation box side-length, if any repeated structures
have small angular separations when projected onto the `mock sky',
then projection-effect artefacts can be introduced into the
catalogue. \cite{Blaizot05} illustrate the effect of these artefacts,
along with possible methods for eliminating them. One method that they
demonstrate to be effective is to apply random sequences of ${\rm
\pi/2}$ rotations and reflections to the replicated boxes so that any
repeated structures are viewed at different orientations and appear as
different structures. The problem with this approach is that, due to
the periodic boundary conditions of the N-body simulation, when tiling
the replications, any transformation besides a translation would add
undesirable discontinuities into the underlying density
field. However, if for example one wishes to extract clustering
statistics, the underlying density field should be preserved. We
therefore choose not to use this method in the construction of our
lightcone catalogues.

\begin{figure}
\includegraphics[width=8.6cm,trim=2cm 1cm 1cm 2.5cm, clip=true]{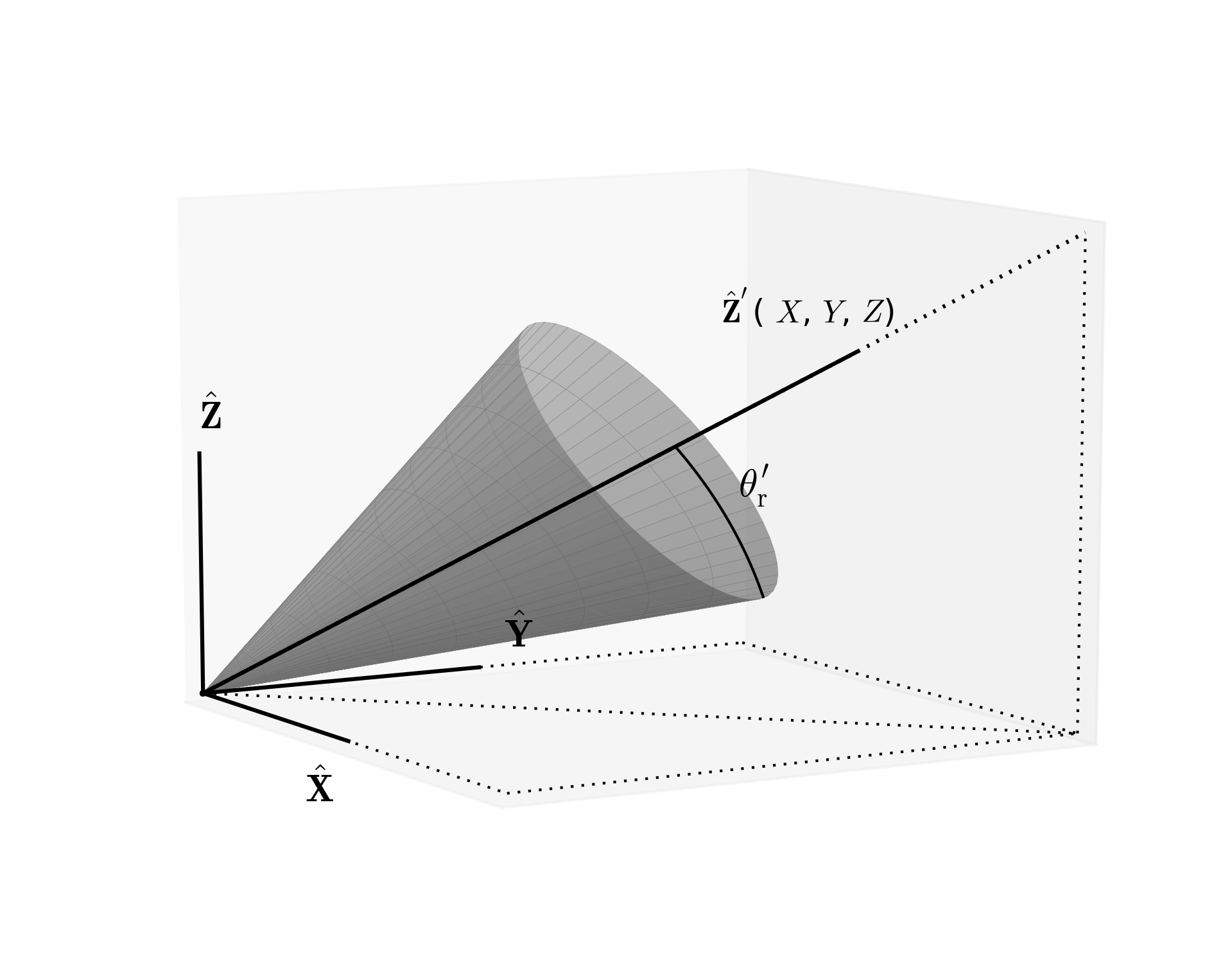}
\caption{Schematic of lightcone geometry. The axis
$\mathbf{\hat{Z}^{\prime}}$ defines the central \emph{line-of-sight}
vector of the observer. The angle $\theta^{\prime}_r$ defines the
angular size of the field-of-view of the lightcone. Any galaxy whose
position vector, $\mathbf{\vec{r}^{\prime}}(X^{\prime}, Y^{\prime},
Z^{\prime})$, is offset from the $\mathbf{\hat{Z}^{\prime}}$ axis by
an angle $\theta^{\prime}>\theta^{\prime}_r$ is excluded from the
lightcone.}
\label{fig:obs_coord_sys}
\end{figure}

\subsection{Orientating the observer}
\label{sec:orienating_observer}

Our aim when orientating the observer is to be able to define a
right-handed Cartesian co-ordinate system, ${\rm
(\mathbf{\hat{X}^{\prime}},\mathbf{\hat{Y}^{\prime}},\mathbf{\hat{Z}^{\prime}})}$,
such that the observer is looking down the $\mathbf{\hat{Z}^{\prime}}$
axis, as illustrated in Fig.~\ref{fig:obs_coord_sys}. This axis
defines the central axis of symmetry of the conical volume of the
lightcone and points to the centre of the field of the lightcone on
the mock sky, i.e. $\mathbf{\hat{Z}^{\prime}}$ points along the
central \emph{line-of-sight} vector of the observer. The half-opening
angle, $\theta^{\prime}_r$, governs the angular extent of the
field-of-view of the lightcone (see
$\S$\ref{sec:finalising_geometry}). The orientation of the observer is
simply how we describe this vector, $\mathbf{\hat{Z}^{\prime}}$, in
terms of the global Cartesian axes of the ``super cube'',
$\mathbf{\hat{Z}^{\prime}}(X,Y,Z)$. 

For deep, pencil-beam mock catalogues, carefully choosing the
orientation of the observer can minimise, or even remove, structure
repetition. The approach adopted by \cite{KW07} is to orientate the
observer in a `slanted' direction, with respect to the Cartesian axes
of the simulation box, so that the observer is not looking along any
of the Cartesian axes or the cube diagonals (along which structure
repetition can introduce noticeable artefacts). By defining the central
line-of-sight of the observer as
$\mathbf{\hat{Z}^{\prime}}(X,Y,Z)=(n,m,nm)$, where $m$ and $n$ are
integers with no common factor, \citeauthor{KW07} are able to
construct lightcone catalogues with a near-rectangular sky coverage of
$1/m^2n \times 1/n^2m$ (radians) in which the first repeated structure
will lie at distance of $\sim mnL_{{\rm box}}$ from the
observer.\footnote{\cite{CW10} adopt a similar approach to
\citeauthor{KW07} by performing volume remapping of the original
simulation box such that the mock catalogue geometry can fit inside
the new geometry without the need for box replication.} When
constructing lightcone catalogues for which we wish to minimise
duplicated structures, albeit at the expense of the solid angle of the
catalogue, we adopt this approach. This is necessary for applications
considering, for example, the angular clustering of galaxies, where
projection effects could severely distort the clustering signal.

Once we are satisfied with the chosen orientation of
$\mathbf{\hat{Z}^{\prime}}$, we define the axis
$\mathbf{\hat{X}^{\prime}}$ (to be perpendicular to both
$\mathbf{\hat{Z}^{\prime}}$ and $\mathbf{\hat{X}}$) and the axis
$\mathbf{\hat{Y}^{\prime}}$ (to be perpendicular to both
$\mathbf{\hat{X}^{\prime}}$ and $\mathbf{\hat{Z}^{\prime}}$).

\subsection{Finalising the lightcone geometry}
\label{sec:finalising_geometry}
Now that we know the location and orientation of the observer we can
set about applying the necessary geometrical cuts to construct the
lightcone volume. The first step is to isolate a spherical volume
about the observer, with a co-moving radius, $r_{{\rm max}}$, whose
value is sufficiently large that, given the flux limits of the survey
we wish to emulate, we would expect to be well into the high-redshift
tail of the galaxy redshift distribution (such that only a negligible
fraction of the brightest, high redshift galaxies are missed). This
radial cut is applied to help speed up the calculation so that we are
not searching for galaxies in box replications that are too far from
the observer to contribute a significant number of objects to the mock
catalogue. For boxes with a fraction of their volume lying within
$r_{{\rm max}}$, we check when each galaxy will enter the observer's
past lightcone (see $\S$\ref{sec:lightcone_selection}). If a galaxy
enters the lightcone at a distance greater than $r_{{\rm max}}$ (or it
never enters the lightcone at all) then it is discarded.

Next we apply an angular cut on the mock galaxies, which is dictated
by the solid angle of the galaxy survey we wish to mimic. The solid
angle in steradians, $\Omega$, of the mock catalogue is defined by
\begin{equation}
\Omega = 2\pi \left [ 1-\cos\left ( \theta^{\prime}_r \right ) \right
 ],
\label{eq:solid_angle}
\end{equation}
where $\theta^{\prime}_r$ is the field-of-view angle of the
catalogue. By varying the value of $\theta^{\prime}_r$ we can
construct lightcones with solid angles ranging from pencil beams, to
all-sky ($2\pi$) surveys.\footnote{By setting $\theta^{\prime}_r=\pi$
we can construct all-sky lightcone catalogues. When constructing such
catalogues we can apply an additional geometrical cut to remove
galaxies that would be obscured by the plane of the Milky Way. Having
calculated the celestial co-ordinates of the galaxy on the mock sky,
we determine the galactic latitude, $b$, of the galaxy and reject all
galaxies with $|b|<b_{{\rm lim}}$, where $b_{{\rm lim}}$ is the
user-specified galactic latitude limit. The solid angle of the all-sky
lightcone is then calculated as, $\Omega_{{\rm all-sky}} = 4\pi -
2\pi\left [ \sin\left ( b_{{\rm lim}} \right ) - \sin\left ( -b_{{\rm
lim}}\right )\right ]$.} Following this cut, the catalogue volume
resembles the sector of a sphere, with half-opening angle
$\theta^{\prime}_r$ and $\mathbf{\hat{Z}^{\prime}}$ as its axis of
symmetry. For those boxes whose volume overlaps that of the catalogue,
we calculate the position at which each galaxy enters the
lightcone. Using this position we calculate the angle
$\theta^{\prime}$, the angle between the position vector of the galaxy
and the $\mathbf{\hat{Z}^{\prime}}$ axis, and discard any galaxy with
$\theta^{\prime}>\theta^{\prime}_r$, that lies outside the solid angle
of the catalogue.

Finally, for those galaxies that are successfully included in the
lightcone, we determine their right-ascension, $\alpha$, and
declination, $\delta$, on the mock `sky'. We do this by first defining
a sky coordinate system such that the observer's central line-of-sight
vector, $\mathbf{\hat{Z}^{\prime}}$, points towards a right ascension,
$\alpha_0$, and declination, $\delta_0$, on the sky. We then determine
the sky position of a galaxy by passing $\mathbf{\vec{r}}$$(X,Y,Z)$,
through the transformation,
$\mathcal{R}$$_Z(\alpha_0)$$\mathcal{R}$$_Y(\pi/2-\delta_0)$
$\mathbf{\vec{r}}$, where $\mathcal{R}$$_Z$ and $\mathcal{R}$$_Y$ are
the standard 3-dimensional Cartesian rotation matrices about the
$\mathbf{\hat{Z}}$ and $\mathbf{\hat{Y}}$-axes respectively. (We
assume that lines of constant declination lie parallel to the $X-Y$
plane, so do not apply any rotation about the $X$ axis).

\subsection{Positioning galaxies within the lightcone}
\label{sec:lightcone_selection}
The lightcone selection of galaxies is carried out by identifying those
galaxies whose light has sufficient time to reach the
observer. However, before one can calculate when a galaxy enters the
lightcone, one must determine the epoch at which its host dark matter
halo enters the observer's past lightcone.

\subsubsection{Placement of halo centres}
\label{sec:halo_placement}
A halo, located at $\mathbf{\vec{r}}(X,Y,Z,t)$, at a lookback time,
$t$, will be ``visible'' to the observer at all instances where $\vert
\mathbf{\vec{r}}(X,Y,Z,t) \vert \leqslant r_{{\rm c}}(t)$, where
$r_{{\rm c}}$ is the maximum distance that light could have travelled
in the time $t$, i.e. the maximum co-moving, radial distance that is
visible to the observer. For a flat cosmology, i.e. with $\Omega_k=0$,
at the epoch corresponding to redshift, $z$, the maximum co-moving,
radial distance, $r_{{\rm c}}$, that is visible to an observer at the
present day is given by,
\begin{equation}
r_{{\rm c}}(z) = {\displaystyle \int^z_0\frac{c\,{\rm
d}z^{\prime}}{H_0\sqrt{\Omega_{\rm m}\left ( 1+z^{\prime} \right )^3 +
\Omega_{{\rm \Lambda}}}}},
\label{eq:r_c(z)}
\end{equation}
where $H_0$ is the Hubble Constant at the present day, $\Omega_{{\rm
m}}$ is the matter density of the Universe and $\Omega_{{\rm
\Lambda}}$ is the vacuum energy density of the Universe but at the
present day. To construct a mock galaxy catalogue, we place each halo
at the epoch at which it enters the observer's past lightcone,
i.e. the epoch at which the halo would first become ``visible'' to the
observer. If this epoch corresponds to a redshift, $z$, then the halo
is placed at the position, $\mathbf{\vec{r}} (X,Y,Z,z)$, at which
\begin{equation}
\vert \mathbf{\vec{r}} (X,Y,Z,z) \vert = r_{{\rm c}}(z).
\label{eq:cross_lightcone}
\end{equation}
 
Each snapshot, $i$, in the \emph{Millennium Simulation} corresponds to
a discrete cosmic epoch, with redshift, $z_i$. To determine when a
halo enters the lightcone, we loop over the simulation snapshots
searching for the time-step during which Eq.~(\ref{eq:cross_lightcone})
is satisfied. By comparing the position,
$\mathbf{\vec{r}}_j(X_j,Y_j,Z_j,z_i)$, of a halo $j$ that exists at
$z_i$ to the maximum co-moving distance, $r_{{\rm c}}(z_i)$, that is
visible at that epoch, and doing the same for the descendant of the
halo, labelled $k$, that exists at the next snapshot, $z_{i+1}<z_i$,
we can determine whether halo $j$ will enter the lightcone between the
snapshots $i$ and $i+1$, i.e. whether $z_{i+1}<z<z_i$.

Using $\mathbf{\vec{r}}_j(X_j,Y_j,Z_j,z_i)$ and
$\mathbf{\vec{r}}_k(X_k,Y_k,Z_k,z_{i+1})$ as boundary conditions, we
interpolate along the orbital path of the halo $j$ to find the exact
epoch, $z$, at which it enters the lightcone and the position,
$\mathbf{\vec{r}}_j(X,Y,Z,z)$, at which this occurs. We use a cubic
polynomial to describe the position of the halo, in each Cartesian
direction, as a function of the time $t$ between the adjacent
snapshots (i.e. $t_{i+1}<t<t_i$). For example, the Cartesian $X$
component of the path is given by,
\begin{equation}
X(t) = A_Xt^3 + B_Xt^2 + C_Xt + D_X,
\label{eq:halo_position}
\end{equation}
where $A_X$, $B_X$, $C_X$ and $D_X$ are coefficients that can be
determined by requiring that the boundary conditions
($X(t=t_i)=X_j(t_i)$, $X(t=t_{i+1})=X_k(t_{i+1})$,
$\dot{X}(t=t_i)=\dot{X}_j(t_i)$,
$\dot{X}(t=t_{i+1})=\dot{X}_k(t_{i+1})$) are satisfied. The $X$
component of the velocity of the galaxy at time, $t$, is given by the
derivative of Eq.(\ref{eq:halo_position}) with respect to
time. Equations similar to Eq.(\ref{eq:halo_position}) can be derived
for the $Y(t)$ and $Z(t)$ components. The centre of mass of the halo
is then placed at $\mathbf{\vec{r}}_j(X,Y,Z,z)$.

Our decision to use interpolation to determine halo positions is an
extension of the method of \cite{KW07}, who explicitly chose not to
use interpolation but instead placed halos according to the snapshot
with the epoch closest to the one at which the halo enters the
lightcone. \citeauthor{KW07} adopted this approach because of the
difficulties inherent in using interpolation to predict realistic
orbital paths for satellite galaxies. In the next section, we discuss
these difficulties and suggest a solution that provides a good
approximation for our purposes.

\subsubsection{Placement of galaxies}
\label{sec:galaxy_placement}

\begin{figure*}
\includegraphics[width=\textwidth]{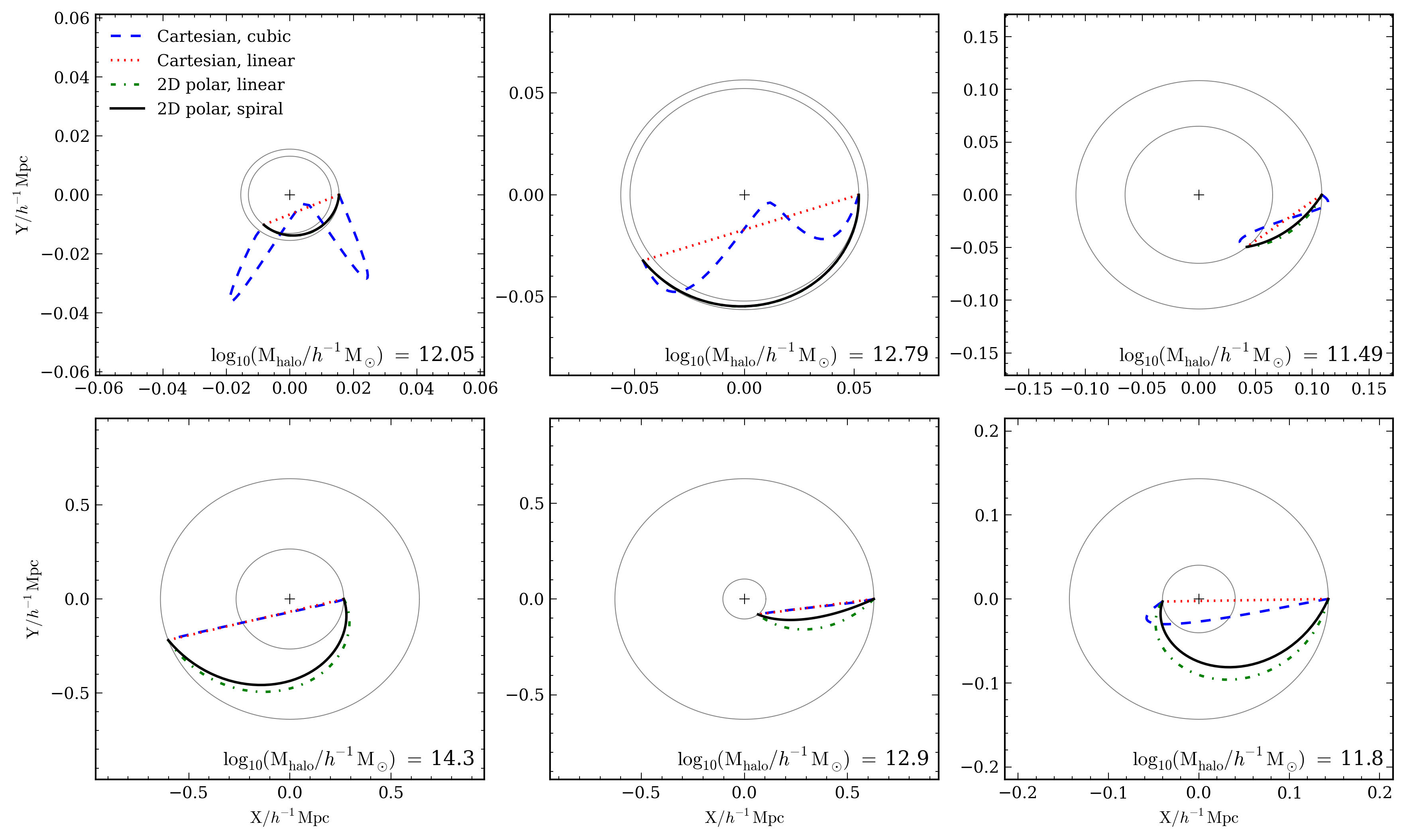}
\caption{Examples demonstrating the modelling of the orbital paths of
satellite galaxies between two adjacent simulation snapshots using
different interpolation schemes. The positions of the satellite
galaxies are displayed relative to the centre of mass of the halo,
which is marked with a $+$. Circles show circular orbits at the start
and end radii of the path of the satellite galaxy. The various
interpolation schemes, which use either 3-dimensional Cartesian
co-ordinates or 2-dimensional polar co-ordinates, are discussed in
$\S$\ref{sec:galaxy_placement} and are shown using different line
colours and styles, as indicated by the key in the top left panel. For
the application presented in $\S$\ref{sec:BzK_section} we use the
2-dimensional polar, linear interpolation scheme.}
\label{fig:sat_interp}
\end{figure*}

The finite spatial extent of halos means that central and satellite
galaxies within a halo will enter the lightcone at slightly different
times. Central galaxies are positioned on the most bound particle of
the most massive \SUBFIND{} group (see
$\S$~\ref{sec:Millennium_trees}) and are at rest relative to the
halo. The location and time at which a central galaxy enters the
lightcone is thus equal to that of its host halo and so for these
galaxies we can use the calculation for the halo centre, as presented
in $\S$\ref{sec:halo_placement}. However, satellite galaxies can enter
the lightcone at an earlier or later epoch than the centre of the host
halo. When positioning a satellite galaxy we can still interpolate
over the evolutionary path of its host halo, but we must first correct
the spatial positions along the path to account for the relative
offset between the position of the satellite galaxy and the centre of
the halo. Therefore, we first need a model to describe the orbital
path of a satellite within its host halo.

\begin{figure*}
\includegraphics[width=\textwidth]{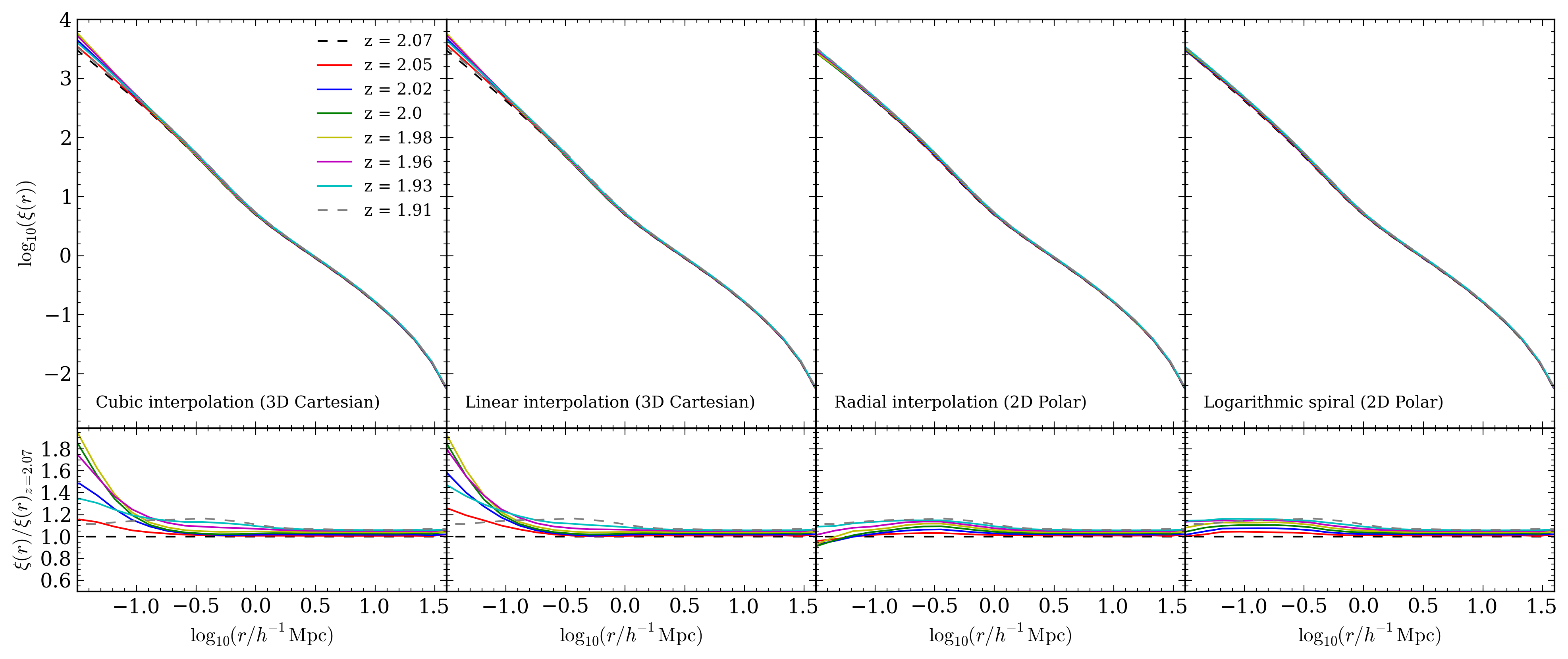}
\caption{The real-space correlation function of galaxies predicted
using four different satellite interpolation schemes: cubic
interpolation in 3D Cartesian space (far left), linear interpolation
in 3D Cartesian space (middle left), radial interpolation in 2D polar
space (middle right) and modelling the satellite orbits using a
logarithmic spiral in 2D polar space (far right). The upper panels
show the correlation function for galaxies at two adjacent simulation
snapshots (corresponding to redshifts $z=1.91$ and $z=2.07$, grey and
black dashed lines) and the same galaxies at six intermediate
redshifts (various solid, coloured lines). The lower panels show the
ratio of each correlation function, relative to the correlation
function measured at the $z=2.07$ snapshot.}
\label{fig:interp_methods_clustering}
\end{figure*}

Modelling physically viable satellite orbits is a non trivial
task. Difficulties arise when the large orbital velocities of
satellite galaxies lead to orbital time-scales that are much shorter
than the spacing of the simulation snapshots. Care must therefore be
taken to ensure that numerical artefacts do not introduce large
positional errors, which might in turn lead to inaccurate predictions
for the one-halo term in the galaxy correlation function.

If, for example, we attempt to describe the orbital path of a
satellite galaxy using a cubic polynomial in Cartesian space that is
constrained to satisfy both the position and velocity boundary
conditions, then in rare instances we may find that the large orbital
velocities of satellite galaxies lead to orbital paths that are highly
eccentric and extend out to large orbital radii. In the majority of
cases where orbital velocities are small, the cubic function fits an
orbital path similar to a simple linear interpolation scheme, which
ignores the velocity boundary conditions. Example orbits, modelled
using different interpolation schemes, are shown in
Fig~\ref{fig:sat_interp}. Unfortunately, if a halo and its descendent
are found on opposing sides of the halo centre of mass, then these
interpolation schemes would lead to satellites being positioned much
closer to the centre of mass of the halo than they should be. This
would have the effect of boosting the clustering signal on small
scales, as shown in the two left-hand panels of
Fig~\ref{fig:interp_methods_clustering}.

Since measurements of galaxy clustering are integral for achieving
many of the goals set by current and future galaxy surveys, we choose
to prioritise the preservation of the galaxy clustering signal in real
space. We do this by moving to a 2-dimensional (2D) plane, defined by
the position of the halo centre of mass and the positions of the
satellite, $j$, and its descendent, $k$, relative to the halo
centre. By assuming that the orbit of satellite $j$ is restricted to
this plane, we use linear interpolation to express the change in the
polar co-ordinates of the orbit of $j$ (relative to the centre of mass
of the halo, located at the origin) as a function of time between the
snapshot epochs, $t_{i+1}<t<t_i$.

We describe the angular position, $\phi$, of the satellite along its
orbit as 
\begin{equation}
\phi(t) = \phi_j(t_i) + \left [ \phi_k(t_{i+1})-\phi_j(t_i) \right ]
\left [ \frac{t-t_i}{t_{i+1}-t_i} \right ].
\label{eq:angular_interp}
\end{equation}
A caveat is that we perform the interpolation along the path that
corresponds to the smallest angular separation between the position of
a satellite and its descendent, which may lead to satellites changing
directions. To describe the change in the radius, $\rho$, of the orbit
of the satellite we can choose to either linearly interpolate the
radius in the same way, i.e.
\begin{equation}
\rho(t) = \rho_j(t_i) + \left [ \rho_k(t_{i+1})-\rho_j(t_i) \right ]
\left [ \frac{t-t_i}{t_{i+1}-t_i} \right ],
\label{eq:radial_interp}
\end{equation}
or couple the radius to the angle using a simple model, such as a
logarithmic spiral,
\begin{equation}
\rho(t) = a\cdot {\rm e}^{b\cdot \phi(t)},
\label{eq:log_spiral}
\end{equation}
where $a=\rho_j(t_i)$ and $b=\phi_k(t_{i+1})\ln \left (
\rho_k(t_{i+1}) / \rho_j(t_i) \right )$. Note that in these two cases
we are ignoring the velocity boundary conditions and assuming that
$\ddot{\rho}(t) = \ddot{\phi}(t) = 0$. However, as can be seen in
Fig~\ref{fig:interp_methods_clustering}, these methods preserve the
galaxy clustering to smaller scales than possible with the
3-dimensional (3D) cubic or linear approaches. Note that in the above
cases, we can interpolate the orbital velocities of satellites using
the same equations as used for the positions. For the application in
$\S$~\ref{sec:BzK_section}, we have adopted to interpolate the
satellite positions using Eq.(\ref{eq:angular_interp}) and
Eq.(\ref{eq:radial_interp}), i.e. a 2-D polar linear interpolation of
both the angle and radius of the satellite orbit.

By converting back to 3D Cartesian coordinates we can express the
epoch, $z$, at which the satellite enters the lightcone as the
position at which,
\begin{equation}
\vert \mathbf{\vec{r}_{\rm halo}}
(X,Y,Z,z)+\mathbf{\vec{r}}^{\prime}_j(X,Y,Z,z) \vert =
r_{{\rm c}}(z),
\label{eq:sat_cross_lightcone}
\end{equation}
where $\mathbf{\vec{r}_{\rm halo}} (X,Y,Z,z)$ is the global position
of the dark matter hosting the satellite at this epoch and
$\mathbf{\vec{r}}^{\prime}_j(X,Y,Z,z)$ is the position of the
satellite relative to the centre of this halo.

Fig.\ref{fig:interp_methods_clustering} shows that between $z=1.91$
and $z=2.07$ (which is a $5\%$ change in $1+z$) there is a $20\%$
difference in the amplitude $\xi(r)$. If we did not interpolate the
position of satellite galaxies between snapshots, then at a redshift
intermediate to $z=1.91$ and $z=2.07$, there would be up to a $\sim
10\%$ error in the correlation function. Whilst any interpolation
scheme is approximate, it is clear that it is better to make an
attempt to adjust the galaxy positions if the lightcone crossing
occurs between snapshots, rather than jumping from one set of fixed
positions to the other, which would result in discontinuities in the
correlation function.

\subsection{Treatment of galaxy properties in the lightcone}
\label{sec:galaxy_properties}

\subsubsection{Intrinsic properties}
\label{sec:intrinsic_properties}
For each galaxy that enters the lightcone and satisfies the
geometrical cuts described in $\S$\ref{sec:finalising_geometry}, we
need to output galaxy properties (i.e. stellar mass, SFR, etc.)
that are appropriate for the epoch at which we have placed the galaxy.

With knowledge of the star-formation history of the galaxy, we can
follow the evolution of any galaxy property over cosmic time. However,
as with galaxy positions, this information is only recorded at the
discrete epochs corresponding to the simulation snapshots. Ideally we
would like to again use interpolation to determine the value for any
galaxy property at any given epoch. Unfortunately the evolution of the
majority of galaxy properties is complex and by interpolating between
the snapshot epochs we risk over-simplifying this evolution and
deriving incorrect values. For instance, the build-up of the stellar
mass of a galaxy between two consecutive snapshots will receive
contributions from many different sources. Besides quiescent star
formation in the disk, many other events, such as disk instabilities
or mergers with one or more other galaxies, can lead to starbursts and
a SFR that is highly variable with time. In the case of galaxy mergers
we cannot accurately say, from the snapshot data alone, when during a
time-step the merger occurred. Therefore interpolation over the
properties of each progenitor may lead to double counting and, at the
epochs at which they enter the lightcone, each progenitor having
properties that are (possibly significantly) mis-estimated.

We could evaluate galaxy properties for any given epoch by solving the
set of coupled differential equations that govern the exchange of
material between the hot gas in the halo and the cold gas and stars in
the galaxy. However, this exercise is non-trivial and would require
the full calculation performed by \GALFORM{} to be reproduced for each
galaxy out to the epoch at which it enters the lightcone, which would
be computationally expensive. Similarly, we could have originally run
\GALFORM{} and output the galaxy properties on a finer time mesh.
However, this would extend the run time of the model and take up
significantly more disk space. Instead, we adopt a procedure similar
to that of \cite{KW07} and, for any galaxy that enters the lightcone,
we assign the galaxy the intrinsic properties it had at the snapshot
immediately prior to the epoch, $z$, at which it entered the
lightcone, i.e. the snapshot $i$ with the smallest redshift, $z_i$,
for which $z_i>z$.

\begin{figure}
\begin{center}
\includegraphics[width=8.6cm]{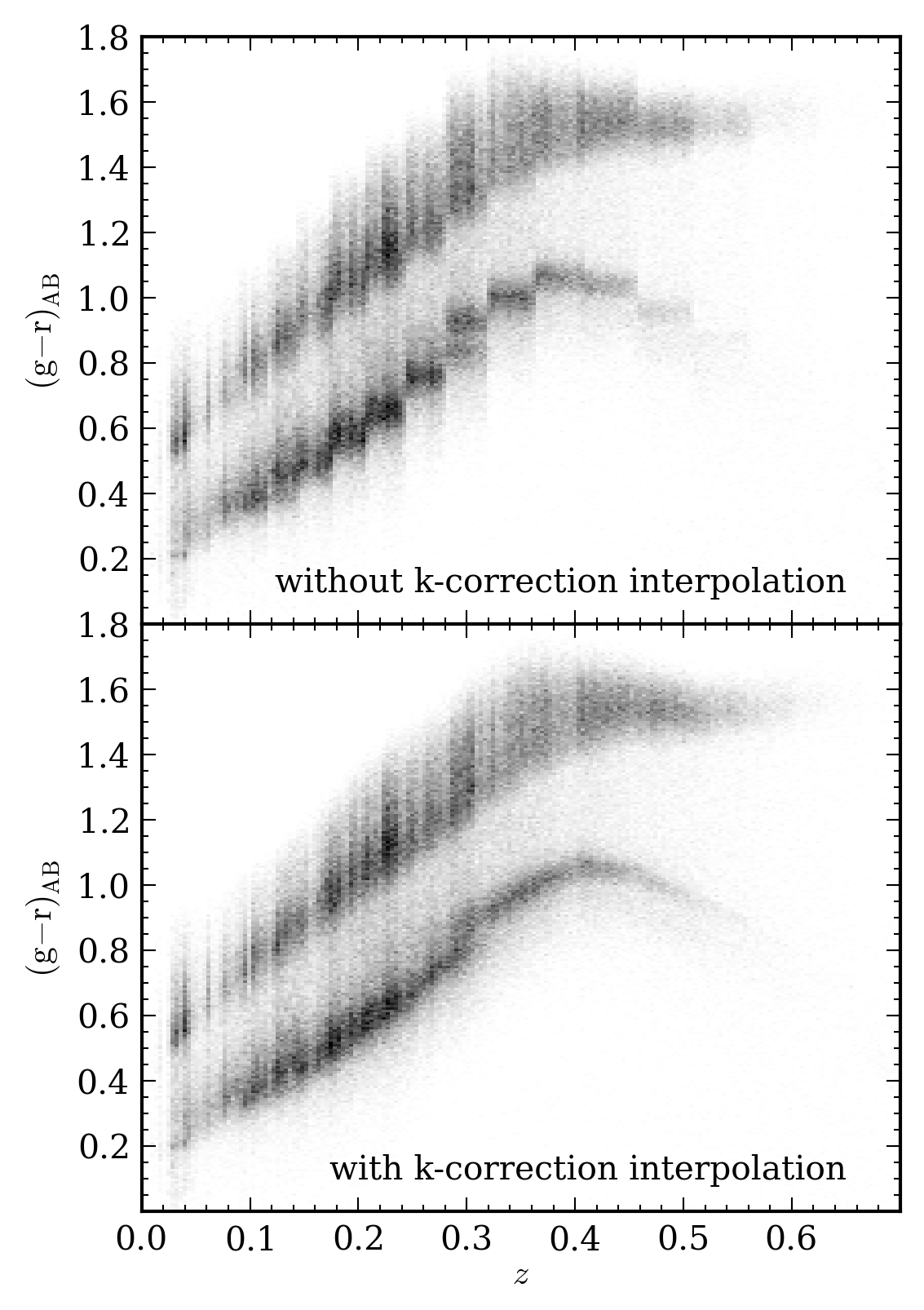}
\caption{Distribution of SDSS $g-r$ colour as a function of redshift,
$z$, in a lightcone catalogue constructed for the GAMA survey, both
without k-correction interpolation (upper panel) and with k-correction
interpolation (lower panel, see $\S$\ref{sec:observed_props} for
details). Shading corresponds to the number density of galaxies. (Note
that the apparent vertical stripes in the galaxy distributions
correspond to peaks in the galaxy redshift distribution.)}
\label{fig:k_interp}
\end{center}
\end{figure}

\begin{figure*}
\begin{center}
\includegraphics[width=\textwidth]{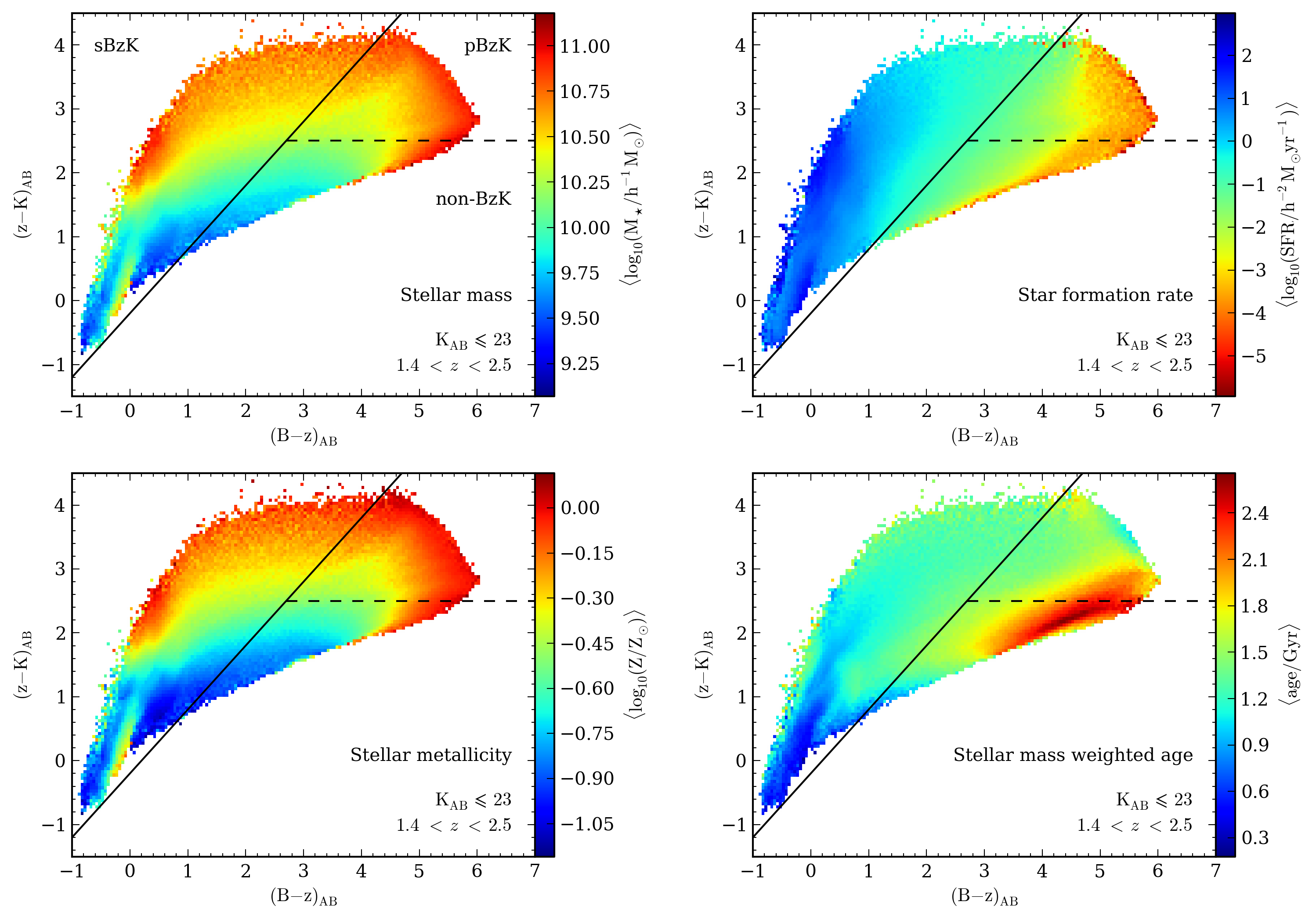}
\caption{The predicted distribution of ${\rm K_{\AB}}\leqslant23$
galaxies with $1.4<z<2.5$ in the BzK colour plane, colour coded, as
indicated by the key on the right of each panel, according to the
median value in a 2-dimensional colour-colour bin for different galaxy
properties: stellar mass (upper left), star-formation rate (upper
right), stellar metallicity (lower left) and stellar mass weighted age
(lower right). The solid and dashed lines correspond to the BzK
criteria used by \protect\cite{Daddi04b} (see ~\ref{sec:BzK_intro} for
further details). The sBzK and pBzK regions have been labelled in the
upper left panel.}
\label{fig:BzK_property_planes}
\end{center}
\end{figure*}

\subsubsection{Observed properties}
\label{sec:observed_props}
Having set the intrinsic properties of the galaxies, we can now use
this information, along with their positions, to evaluate their
observed properties, namely their observed fluxes (and apparent
magnitudes). At this point we need to use the position of a galaxy to
derive its luminosity distance, $d_L$, which is required to relate the
emitted luminosity per unit frequency, $L_{\nu}(\nu_{{\rm e}})$, of an
object to its observed flux per unit frequency, $S_{\nu}(\nu_{{\rm
o}})$. For a flat Universe the luminosity distance out to a redshift
$z$ is simply, $d_L(z) = r_c(z)(1+z)$. Therefore a galaxy in the
lightcone at a cosmological redshift $z$ will have an observed flux,
\begin{equation}
S_{\nu}(\nu_{{\rm o}}) = \left ( 1+z \right ) \frac{L_{\nu}(\nu_{{\rm
e}})}{4\pi d_L^2\left (z\right )},
\label{eq:lum_to_flux}
\end{equation}
where $\nu_{{\rm o }}$ is the observed frequency of the light from the
galaxy. The emitted (rest-frame) frequency is related to the observed
(observer-frame) frequency by $\nu_{{\rm e}} = \nu_{{\rm o}}\left (
1+z\right )$. The observer-frame apparent magnitude of a galaxy, in
the AB system, is then given by,
\begin{equation}
m_{\AB} = -2.5\log_{10}\left [ \frac{\int S_{\nu}(\nu_{{\rm
o}})R(\nu_{{\rm o}}){\rm d}\nu_{{\rm o}}}{S_{\nu_{{\rm o}}}\int
R(\nu_{{\rm o}}){\rm d}\nu_{{\rm o}}} \right ],
\label{eq:app_mag_from_flux}
\end{equation}
where $R(\nu_{{\rm o}})$ is the filter response of a specified
photometric band and $S_{\nu_{{\rm o}}}$ is the AB reference flux per
unit frequency \citep{OG83}.

In our case, \GALFORM{} calculates the emitted luminosity of a galaxy,
so we can calculate the observer-frame absolute magnitude, $M_{\AB}$,
of the galaxy (assuming $z\ne 0$) from,
\begin{equation}
M_{\AB} = -2.5\log_{10}\left [ \frac{\int L_{\nu}(\nu_{{\rm
e}})R(\frac{\nu_{{\rm e}}}{1+z}){\rm d}\nu_{{\rm e}}}{L_{\nu_{{\rm
o}}}\int R(\frac{\nu_{{\rm e}}}{1+z}){\rm d}\nu_{{\rm e}}} \right ],
\label{eq:app_mag_from_flux}
\end{equation}
where $L_{\nu_{{\rm o}}}$ is now the AB reference luminosity,
$L_{\nu_{{\rm o}}} = 4\pi(10{\rm pc})^2S_{\nu_{{\rm o}}}$. From this
we can calculate the observer-frame apparent magnitude of a galaxy, in
the AB system, by,
\begin{equation}
m_{\AB} = M_{\AB} + 5\log_{10}\left ( \frac{d_L\left ( z\right
)}{10{\rm pc}} \right ) - 2.5\log_{10}\left ( 1+z\right ).
\label{eq:bcdm}
\end{equation}

Due to the large number of galaxies modelled by \GALFORM{}, the full
SED of each galaxy is not stored. Instead, the luminosity is computed
in a set of filters specified at run time. Hence the definition of the
filter response in the galaxy rest frame, $R(\nu_{{\rm e}}/(1+z))$, is
tied to the output redshifts of the simulation snapshots and the
k-correction applied does not correspond to the redshift of the galaxy
in the lightcone. This discrepancy leads to visible discontinuities in
distributions involving the photometric properties of the galaxies,
such as galaxy colours versus redshift, as shown in the upper panel of
Fig.~\ref{fig:k_interp}. The breaks apparent in the distribution
correspond to the redshifts of the simulation snapshots.

As discussed in the previous section, the complex time dependence of
galaxy luminosity means that we cannot simply interpolate the absolute
magnitudes. However, since the size of the wavelength shift applied to
a filter depends only on the redshift to the galaxy (and not on any of
its intrinsic properties), then we can apply a correction to all the
observer-frame absolute magnitudes (and dust emission luminosities) to
take into account the redshift of lightcone crossing. Consider again a
galaxy, $j$, that is originally found in the snapshot, $i$, at
redshift $z_i$, and which has a descendent in snapshot ${i+1}$, at a
redshift $z_{i+1}<z_i$. Assume that the galaxy has an observer-frame
absolute magnitude $M_j(z_{i})$. Since the wavelength shift applied
depends only on the redshift of a galaxy, we can easily predict the
observer-frame absolute magnitude that $j$ would have if placed at the
redshift of its descendent in snapshot $i+1$, i.e. $M_j(z_{i+1})$
within the \GALFORM{} code, but with a star-formation history computed
up to $t_i$. If the galaxy, $j$, enters the lightcone at an
intermediate epoch, $z$, then we can interpolate between $M_j(z_i)$
and $M_j(z_{i+1})$ to estimate $M_j(z)$. Note that by interpolating
the magnitudes (and luminosities) in this way, we have not changed the
shape of the SED of the galaxy, but rather have applied a further
systematic wavelength shift to the galaxy SED. As can be seen in the
lower panel of Fig.~\ref{fig:k_interp}, this correction, which was
also applied by \cite{Blaizot05} and \cite{KW07}, smooths out the
`saw-tooth' pattern seen in the colour distribution.

We can also calculate an observed redshift for the mock galaxies,
emulating the measurement that would be taken from a galaxy spectrum
using one or more identified emission lines. The observed redshift,
$z_{{\rm obs}}$, of a galaxy, which includes the cosmological redshift
due to the Hubble flow as well as a component due to the local
peculiar motion of the galaxy, is defined by,
\begin{equation}
z_{{\rm obs}} = \left ( 1+z \right ) \left ( 1+\frac{v_{{\rm
      r}}}{c}\right ) -1,
\label{eq:z_obs}
\end{equation}
where $z$ is the cosmological redshift at which the galaxy enters the
lightcone and $v_{{\rm r}}$ is the radial component of the peculiar
velocity vector, $\mathbf{\vec{v}}$, of the galaxy (i.e. $v_{{\rm
r}}=\mathbf{\vec{v}}\cdot\mathbf{\hat{r}}$, where $\mathbf{\hat{r}}$
is the normalised line-of-sight position vector of the galaxy). We do
not at present include any calculation of photometric redshifts, or
their uncertainties, in \GALFORM{} or our lightcone code. These
properties can be readily calculated in post-processing using the
photometry output for each galaxy.

\subsection{Applying the survey criteria}

The final stage in constructing a mock catalogue is to apply the
radial selection criteria of the survey being mimicked and reject
those galaxies fainter than the specified flux limits. For many
surveys this involves placing a cut on the flux at one or more
wavelengths or an apparent magnitude limit in one or more photometric
bands. We can select galaxies according to any intrinsic or observed
galaxy property. For example, besides generating flux limited
lightcone catalogues, we are able to construct catalogues limited by
stellar mass, atomic hydrogen mass or even halo mass. Given a list of
selection criteria, we can control whether a galaxy must pass just one
or all of these criteria simultaneously in order to be included in the
final catalogue. Note that the lightcone catalogues that we produce
correspond to ideal surveys, i.e. we do not apply any completeness
masks or simulate the loss of galaxies due to poor observing
conditions, fibre collisions or quality of spectra.  Such completeness
effects are survey specific and can be applied to the catalogues in
post-processing.


\begin{figure}
\begin{center}
\includegraphics[width=8.6cm]{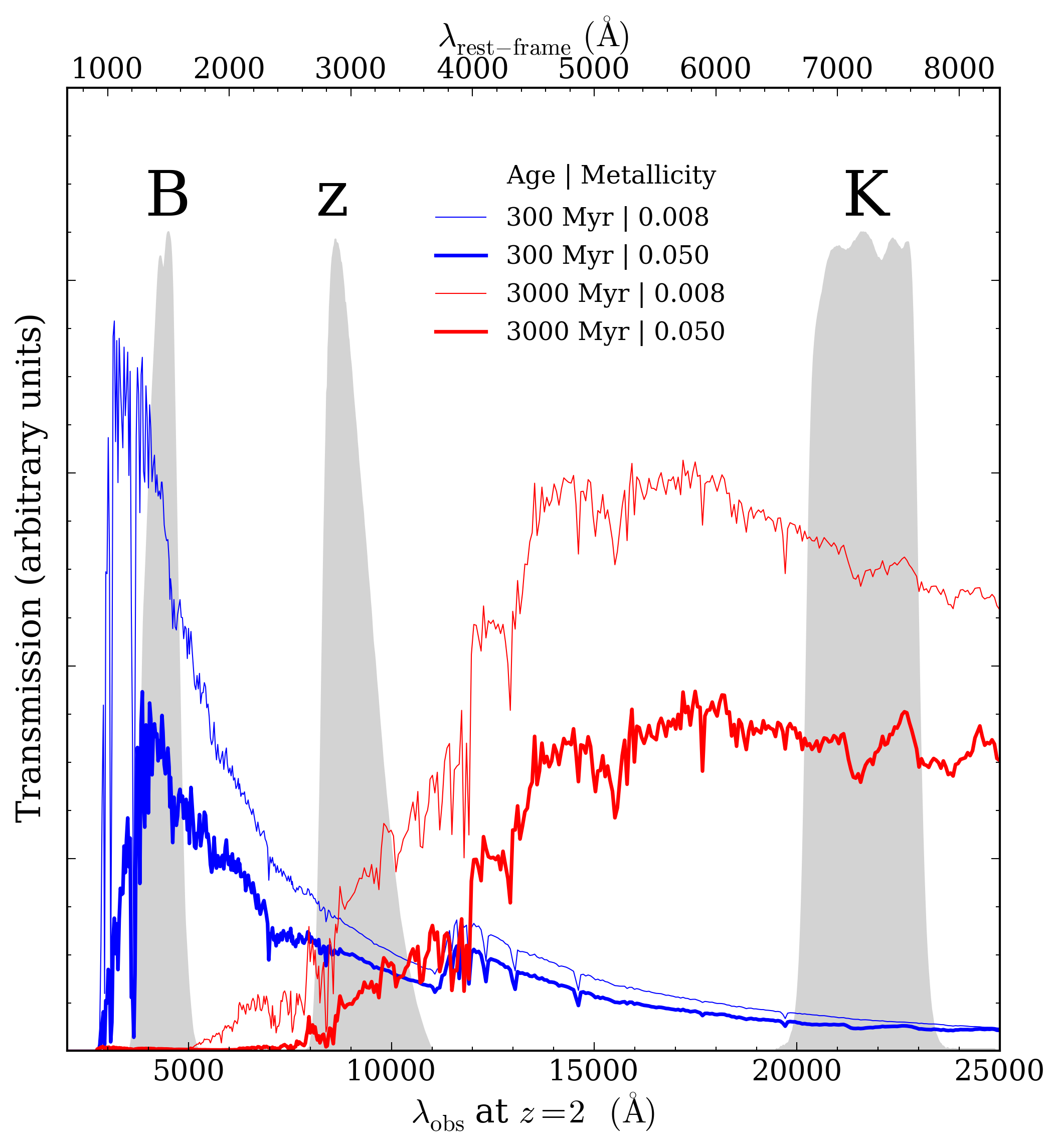}
\caption
{The shading shows the transmission curves of the ${\rm B}$, ${\rm z}$
and ${\rm K}$ filters used by \protect\cite{Daddi04b}. Also shown are
the synthetic spectra (plotted as luminosity per unit wavelength) for
two galaxies at $z=2$. The spectra were obtained using {\tt PEGASE.2}
\protect\citep{PEGASE2}, assuming a \protect\cite{Kennicutt83} IMF,
and a single instantaneous burst of star formation. The spectra are
shown for two different epochs, when the stellar population has an age
of ${\rm 300\,Myr}$ (blue) and an age of ${\rm 3000\,Myr}$ (red) and
for a sub-solar (thin line) and a super-solar (thick line)
metallicity. The flux and transmission units are arbitrary and the
spectra have been normalised so as to be visible on similar scales to
the transmission curves.}
\label{fig:spectraBzK}
\end{center}
\end{figure}

\section{Application to the B\lowercase{z}K colour selection}
\label{sec:BzK_section}
In this section we use a lightcone mock catalogue of $50\,{\rm
deg}^2$, built using the \cite{Bower06} \GALFORM{} model, to study the
properties of galaxies selected using the BzK colour technique. We
have constructed the lightcone by selecting all galaxies with ${\rm
K_{\AB}}\leqslant 24$.

\subsection{The BzK selection}
\label{sec:BzK_intro}
The BzK colour selection is designed to identify galaxies in the
redshift interval $1.4<z<2.5$ based upon their location in the $({\rm
B-z})$ vs. $({\rm z-K})$ colour plane \citep{Daddi04b}. Additionally,
the selection is also advertised as being able to separate
star-forming galaxies from those that are passively evolving.

\citeauthor{Daddi04b} identified star-forming galaxies at $z>1.4$
(referred to as star-forming BzK, or sBzK galaxies) using the
criterion
\begin{equation}
{\rm BzK \geqslant -0.2,}
\label{eq:sBzK}
\end{equation}
where ${\rm BzK \equiv (z-K)_{\AB} - (B-z)_{\AB}}$. This condition is
indicated by the solid black line in
Fig.~\ref{fig:BzK_property_planes}. The sBzK region lies above this
line, as labelled in the upper left panel of
Fig.~\ref{fig:BzK_property_planes}.

To select galaxies at $z>1.4$ that are passively evolving (referred to
as passive BzK, or pBzK galaxies)\footnote{We shall refer to the
combined sBzK and pBzK galaxy population as BzK galaxies.},
\citeauthor{Daddi04b} proposed applying the following conditions:
\begin{equation}
{\rm BzK < -0.2 \,\, {\rm and} \,\, (z-K)_{\AB} > 2.5.}
\label{eq:pBzK}
\end{equation}
The pBzK galaxies populate the region between the solid and dashed
lines in Fig.~\ref{fig:BzK_property_planes}, i.e. the upper right
region of the $({\rm B-z})$ vs. $({\rm z-K})$ colour plane.

The BzK selection works by using colours that sample key features in
the spectral energy distributions of galaxies at $1.4<z<2.5$, mainly
the rest-frame $4000{\rm \AA}$ break and the UV continuum
slope. Fig.~\ref{fig:spectraBzK} shows the synthetic spectra for two
galaxies at $z=2$, obtained using the {\tt PEGASE.2} stellar
population synthesis code \citep{PEGASE2}. The red lines correspond to
a galaxy that is dominated by an old stellar population and thus
exhibits a prominent break around $4000{\rm \AA}$ (in the rest frame),
which is created by the build-up of the absorption lines of ionised
metals. Between $1.4<z<2.5$ the break moves over the observed
wavelength range $\sim9000-15000{\rm \AA}$, between the ${\rm z}$- and
${\rm K}$-bands. In this redshift range, we find that the ${\rm
(z-K)}$ colours of model galaxies become monotonically redder with
increasing strength of the $4000{\rm \AA}$ break.

At $z\leqslant1.4$, the continuum longwards of the $4000{\rm \AA}$
break is shifted into the ${\rm z}$-band, resulting in galaxies at
these redshifts having bluer ${\rm (z-K)}$ colours (and, for a fixed
${\rm B}$-band flux, redder ${\rm (B-z)}$ colours). The BzK criteria
are therefore designed to exclude these galaxies, which lie to the
right of the solid line and below the dashed line in
Fig.~\ref{fig:BzK_property_planes}. However, as we will see in
$\S$~\ref{sec:efficiency_of_BzK}, the finite width of both the break
and the ${\rm z}$-band filter, mean that we would expect some
contamination to occur, as well as the loss of some galaxies within
the target redshift interval, $1.4<z<2.5$.

As can be seen in Fig.~\ref{fig:spectraBzK}, the $4000{\rm \AA}$ break
is stronger for galaxies with old stellar populations or high stellar
metallicity \citep{Kauffmann03a,Kriek06a,Kriek11}. As such, we would
expect old, metal rich galaxies to display redder ${\rm (z-K)}$
colours. The lower two panels in Fig.~\ref{fig:BzK_property_planes}
show the predicted variation of metallicity (left) and stellar mass
weighted age (right) within the ${\rm (B-z)}$ vs. ${\rm (z-K)}$
plane. The predicted trends with ${\rm (z-K)}$ colour agree with the
expectations for the variation of the $4000{\rm \AA}$ break with both
age and metallicity. However, these trends are weakened by the effect
of dust in young galaxies, which reddens the ${\rm (z-K)}$ colour.

To isolate young, metal poor galaxies at $1.4<z<2.5$ that have not yet
developed a strong $4000{\rm \AA}$ break, another spectral feature is
required. Young, star-forming galaxies have steep UV continua due to
the presence of bright, young, hot stars. At $z\sim2$ the UV continuum
is shifted into the optical, as shown in
Fig.~\ref{fig:spectraBzK}. The presence of the steep UV slope boosts
the ${\rm B}$-band flux of these galaxies, leading them to have very
blue ${\rm (B-z)}$ colours, as can be seen in the lower right panel of
Fig.~\ref{fig:BzK_property_planes}. The correlation between UV
luminosity, due to young stars, and star-formation rate, SFR, suggests
that we would also expect a correlation between the SFR of a galaxy
and its ${\rm (B-z)}$ colour for a given ${\rm K}$-band limit. Such a
trend is clearly visible in the upper right panel of
Fig.~\ref{fig:BzK_property_planes}.

\subsection{Predicted numbers of BzK galaxies}
\label{sec:predicted_numbers}

\begin{figure*}
\begin{center}
\includegraphics[trim=2cm 0cm 0cm 0cm,clip=true,width=17.2cm]{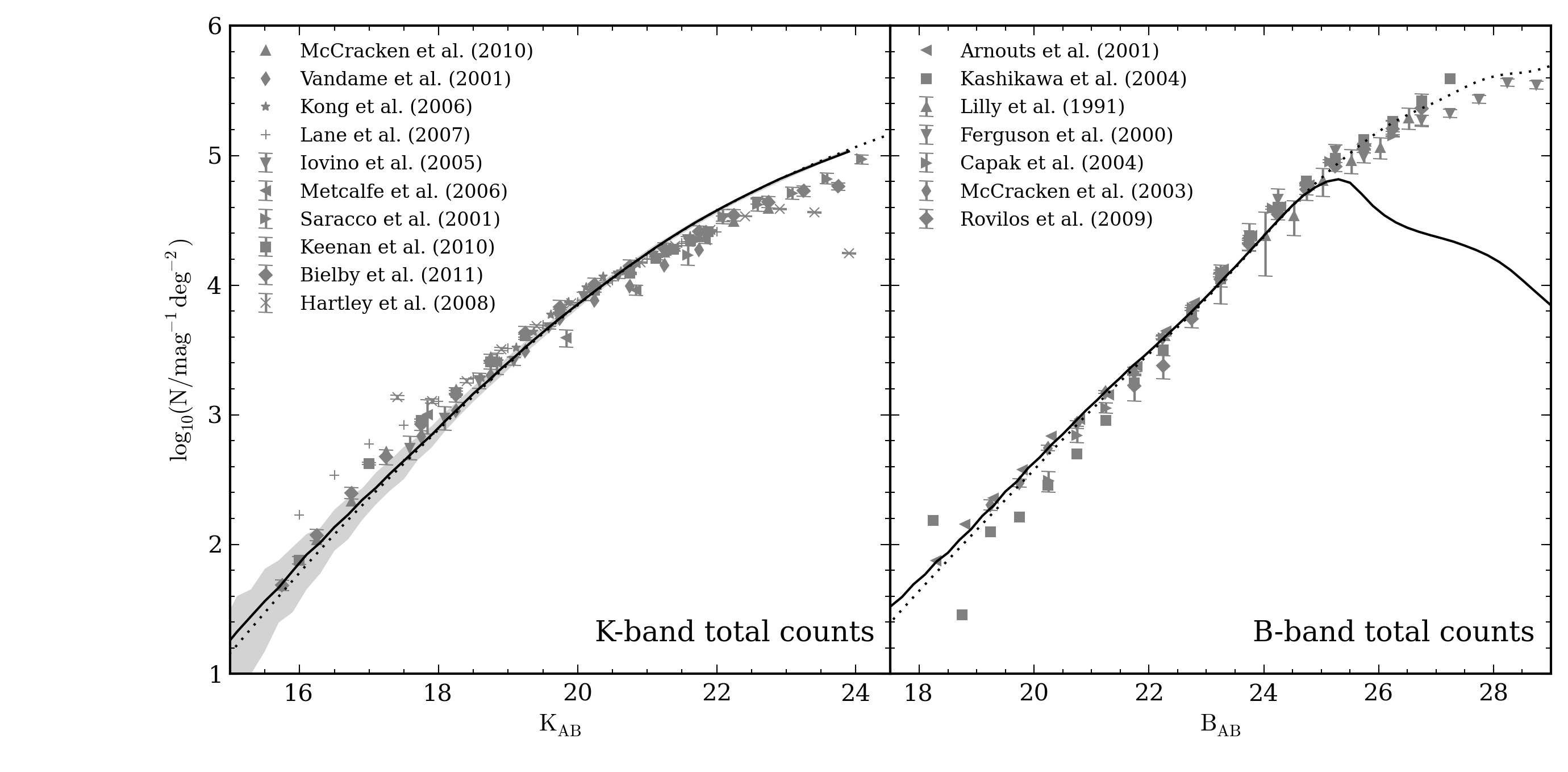}
\caption{Predicted ${\rm K_{\AB}}$-band (left) and ${\rm
    B_{\AB}}$-band (right) differential number counts for all ${\rm
    K_{\AB}}\leqslant 24$ selected galaxies in the lightcone
  constructed using the \protect\cite{Bower06} model (solid
  lines). The dotted lines show the number counts calculated by
  integrating the \GALFORM{} galaxy luminosity function over co-moving
  volume. The latter uses a single band limit, hence the discrepancy
  with the counts from the lightcone in the ${\rm B}$-band. Also shown
  are observationally estimated ${\rm K}$-band number counts from
  \protect\cite{Saracco01,Vandame01,Iovino05,Metcalfe06,Kong06,Lane07,Hartley08,Keenan10,McCracken10,Bielby11*}
  and ${\rm B}$-band number counts from
  \protect\cite{Lilly91,Ferguson00,Arnouts01,McCracken03,Kashikawa04,Capak04,Rovilos09}.
  In the left-hand panel, the light grey shaded region shows the
  $10-90$ percentile spread in the ${\rm K_{\AB}}$-band differential
  number counts for 100 separate $1\,{\rm deg}^2$ lightcone fields.}
\label{fig:total_numbercounts}
\end{center}
\end{figure*}

\begin{figure*}
\begin{center}
\includegraphics[trim=4cm 0.25cm 0cm 0cm,clip=true,width=\textwidth]{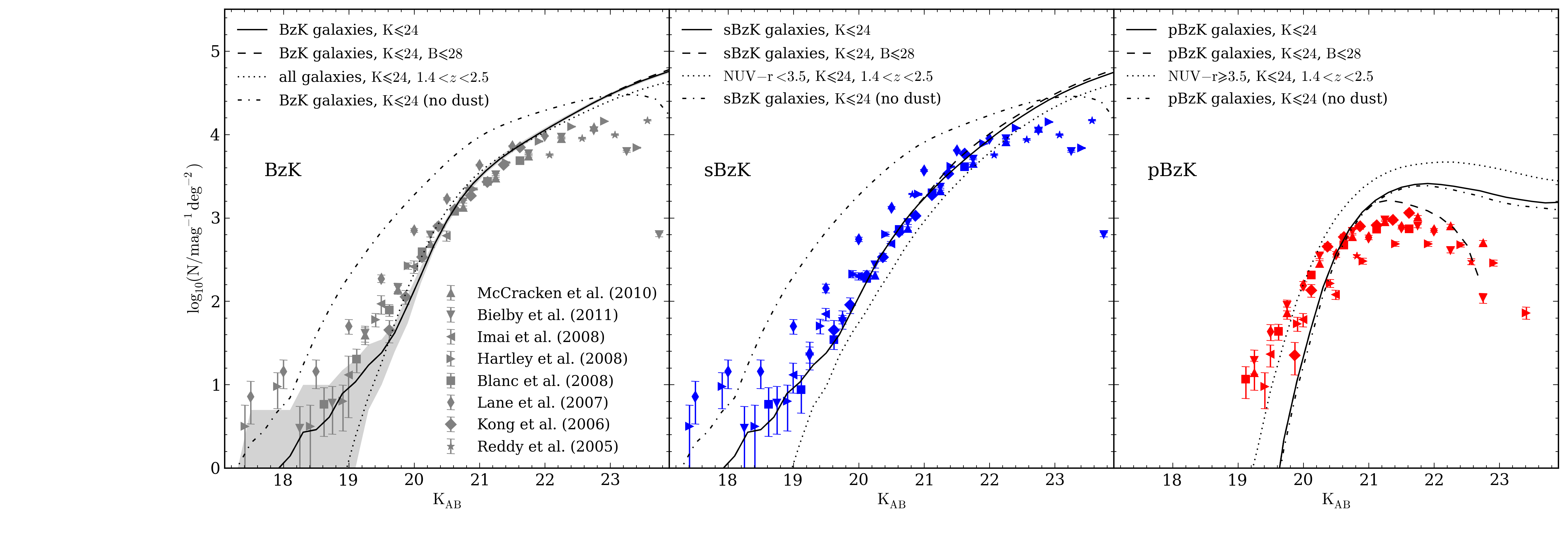}
\caption{Predicted $K_{\AB}$-band differential number counts for all
  BzK (left), sBzK (middle) and pBzK selected galaxies (right) with
  $K_{\AB}\leqslant24$ in the lightcone catalogue (solid lines). The
  dashed lines show the predicted number counts when a ${\rm B}$-band
  detection limit of ${\rm B_{\AB}}\leqslant28$ is considered in
  addition to the ${\rm K}$-band limit (see
  $\S$~\ref{sec:predicted_numbers}, in the left-hand panel, this line
  is underneath the solid one). The dot-dashed lines show the BzK
  number counts when extinction due to dust is omitted (see
  $\S$\ref{sec:dust}). In the left-hand panel, the dotted line shows
  the counts for all $K_{\AB}\leqslant24$ selected galaxies within
  $1.4<z<2.5$. In the middle panel the dotted line corresponds to the
  counts of galaxies with ${\rm NUV-r}<3.5$, $K_{\AB}\leqslant24$ and
  $1.4<z<2.5$ and in the right-hand panel the dotted line corresponds
  to the counts of galaxies with ${\rm NUV-r}\geqslant 3.5$,
  $K_{\AB}\leqslant24$ and $1.4<z<2.5$ (see
  $\S$~\ref{sec:predicted_numbers} for further details on the colour
  cut). Also shown are observationally estimated number counts from
  \protect\cite{Reddy05,Kong06,Lane07,Blanc08,Hartley08,Imai08,McCracken10,Bielby11*}.
  In the left-hand panel, the light grey shaded region shows the
  $10-90$ percentile spread in the ${\rm K_{\AB}}$-band differential
  number counts of BzK galaxies in 100 separate $1\,{\rm deg}^2$
  lightcone fields.}
\label{fig:BzK_numbercounts}
\end{center}
\end{figure*}

Since the BzK technique is used to select a subsample of ${\rm
K}$-band (or ${\rm B}$-band) selected galaxies, we first inspect the
predicted total number counts of all galaxies in the ${\rm B}$ and
${\rm K}$-bands in the mock catalogue, which are shown by the solid
lines in Fig.~\ref{fig:total_numbercounts}. We remind the reader that
our mock catalogue has a solid angle of $50 \,{\rm deg}^2$ and that
galaxies were selected with ${\rm K}_{\AB}\leqslant24$. In both bands,
the \GALFORM{} mock catalogue provides a reasonable match to the
observed counts. The ${\rm B}$-band counts are in excellent agreement
with the observed numbers, though turn over at ${\rm B}_{\AB}\sim 25$
due to the ${\rm K}_{\AB}\leqslant24$ selection used to construct the
lightcone.

As a sanity check, the dotted lines in
Fig.~\ref{fig:total_numbercounts} show the differential number counts
obtained by integrating the \GALFORM{} galaxy luminosity function over
co-moving volume. The excellent agreement between the counts computed
using the luminosity functions directly from the snapshot outputs and
those from the lightcone demonstrates the success of the magnitude
interpolation scheme used to create the lightcone. The light grey
shaded region in Fig~\ref{fig:total_numbercounts} shows the spread (10
to 90 percentile range) in the counts for 100 separate fields each
with a solid angle of $1\,{\rm deg}^2$, a solid angle typical for the
observational datasets we are comparing. These fields were generated
by randomly selecting field centres within the foot print of the
lightcone, with a buffer zone to avoid placing field centres too close
to the edge of the foot print. In the right-hand panel of
Fig.~\ref{fig:total_numbercounts}, the counts computed from the
luminosity function diverge from the predicted counts in the lightcone
due to the ${\rm K}$-band limit used to construct the lightcone
(whereas the integral over the luminosity function is independent of
this limit).

In the left-hand panel of Fig.~\ref{fig:BzK_numbercounts} we show the
number counts of all BzK galaxies with ${\rm K_{\AB}}\leqslant 24$
from the mock catalogue (solid line). Overall the mock provides a
reasonable match to the observed counts. At faint magnitudes (${\rm
  K_{AB}\gtrsim 22}$), the turnover in the observed BzK counts is
sharper than predicted. However, in this region the observations could
be incomplete. The closest agreement between the predictions and
observations occurs for ${\rm K_{AB}\sim 20.5-22.0}$, where there is a
clear change of slope in both the observations and the \GALFORM{}
predictions. At ${\rm K_{AB}\sim21}$, $\sim1/6$ of both observed and
predicted ${\rm K}$-band selected galaxies are also BzK
galaxies. Brightwards of ${\rm K_{AB}}\sim 19.5$, the predicted BzK
counts exceed the counts for ${\rm K}$-band selected galaxies within
$1.4<z<2.5$ (shown by the dotted line) due to low redshift interlopers
(see $\S$~\ref{sec:contamination}). In
Fig.~\ref{fig:BzK_numbercounts}, the light grey shaded region again
shows the $10-90$ percentile spread in the differential number counts
of BzK selected galaxies in 100 separate fields, each with a solid
angle of $1\,{\rm deg}^2$. At bright magnitudes, the extent of this
shaded region indicates that the spread in the observed counts can be
explained as sampling variance arising from the small solid angles
probed.

We now consider the predicted number counts for the subsamples of sBzK
and pBzK galaxies, shown in the middle and right panels of
Fig.~\ref{fig:BzK_numbercounts}. For faint fluxes (${\rm
K_{\AB}}\gtrsim 21$), sBzK galaxies, both observed and predicted,
dominate the BzK population due to the turnover in the pBzK counts
that can be clearly seen in the right-hand panel of
Fig.~\ref{fig:BzK_numbercounts}.

The \GALFORM{} sBzK number counts show a good overall agreement with
the observations. However, the model over-predicts the number of the
faintest sBzK galaxies. This may simply be the result of the observed
sBzK counts becoming incomplete at faint magnitudes. The predicted
number counts of pBzK galaxies are in reasonable agreement with
observations in the range $19.8\lesssim {\rm K_{\AB}} \lesssim
20.8$. However, the model under-predicts the number of brighter pBzK
galaxies and over-predicts the number of fainter galaxies. This
mismatch between semi-analytical predictions and observed pBzK number
counts has previously been shown by \cite{McCracken10}, who compared
their observational counts to the predictions of the mock catalogues
of \cite{KW07}. Moreover, the \citeauthor{KW07} model gave a poorer
match to the observed sBzK counts than we find.

The turnover at faint magnitudes in the counts of pBzK selected
galaxies has been reported by several authors
\citep[e.g. ][]{Lane07,Hartley08,McCracken10,Bielby11*}. The model
also displays a turnover in the pBzK counts, but at ${\rm K_{\AB}}\sim
21$, $\sim 1$ mag fainter than in the data. Both \cite{Hartley08} and
\cite{McCracken10} propose that limited ${\rm B}$-band photometry is
responsible for the turnover. \citeauthor{Hartley08} showed that
reclassifying $\sim34$ per cent of their sBzK galaxies as pBzK
galaxies, would be sufficient to remove the turnover (we will return
to this point in $\S$~\ref{sec:Bband_depth}). We demonstrate the
impact of the depth of the B-band photometry by recalculating the
predicted counts of BzK, sBzK and pBzK galaxies assuming a ${\rm
B}$-band detection limit of ${\rm B_{AB}=28}$ in addition to the ${\rm
K}$-band limit of ${\rm K_{\AB}}\leqslant 24$. Any galaxy in the mock
catalogue with a ${\rm B}$-band magnitude fainter than this is assumed
to be undetected in ${\rm B}$ and its ${\rm (B-z)}$ colour is
calculated assuming ${\rm B_{\AB}}=28$. This is the approach typically
used in observational catalogues to estimate the colours of objects
that are undetected in a band. The effect this has on the counts is
shown by the dashed lines in Fig.~\ref{fig:BzK_numbercounts}. Although
the predicted pBzK counts are still not in full agreement with the
data, applying a ${\rm B}$-band limit has reduced the mismatch. With
the ${\rm B}$-band limit applied, $\sim 50$ per cent of the pBzK
galaxies are relabelled as sBzK galaxies preferentially at the
faintest ${\rm K}$-band magnitudes. This supports the conclusion that
shallow ${\rm B}$-band photometry contributes to the
turnover. \citeauthor{Hartley08} and \citeauthor{McCracken10} both
observed the turnover for ${\rm K}$-band limited samples down to ${\rm
K_{AB}\leqslant 23.5}$ and ${\rm K_{AB}\leqslant 23}$ respectively,
with ${\rm B}$-band detection limits of ${\rm B_{AB,\,lim}=28.4}$ and
${\rm B_{AB,\,lim}=29.1}$ respectively. The predicted excess of faint
pBzK galaxies is also partially a result of the \cite{Bower06} model
predicting too many red galaxies. A substantial fraction of pBzK
galaxies are satellites. These galaxies could be too red due to the
treatment of gas stripping in satellite subhalos \cite[see
][]{Font08}. If we plot the predicted pBzK number counts considering
only central galaxies (without applying any ${\rm B}$-band detection
limit), we find that the predicted excess of faint pBzK galaxies is
reduced, leading to excellent agreement with the observed counts.

The BzK criteria is not the only technique used observationally to
classify galaxies as star-forming at $z\sim 2$. The rest-frame
near-UV/optical colour, ${\rm NUV-r}$, can also be used to separate
star-forming and passive galaxies. Following \cite{Ilbert10}, we
divide the ${\rm K}$-band selected \GALFORM{} galaxies lying within
$1.4<z<2.5$, into star-forming galaxies (i.e. those with ${\rm
  NUV-r}<3.5$) and passively evolving galaxies (${\rm NUV-r}\geqslant
3.5$) and calculate their number counts. As shown in the middle panel
of Fig.~\ref{fig:BzK_numbercounts}, the predicted sBzK counts are
consistently somewhat higher than those predicted for galaxies with a
blue ${\rm NUV-r}$ colour in $1.4<z<2.5$. We note, however, that
low-redshift contamination will exaggerate the counts in the brightest
bins. The predicted number counts of pBzK galaxies are systematically
below the counts of passive galaxies estimated using the ${\rm NUV-r}$
colour. This highlights the sensitivity of the separation of galaxies
into star-forming and passively evolving classes to the precise colour
criteria used.

The mock catalogue is able to reproduce the combined number counts of
all BzK galaxies, as well as providing reasonable agreement with the
counts of ${\rm K}$-band selected galaxies within
$1.4<z<2.5$. Although the model is able to reproduce the predicted
number counts of sBzK galaxies (which dominate the BzK population), it
is unable to reproduce the predicted number counts of pBzK galaxies.

\subsection{Predicted redshift distribution of BzK galaxies}
\label{sec:BzK_zdist}

\begin{figure}
\begin{center}
\includegraphics[width=8.6cm]{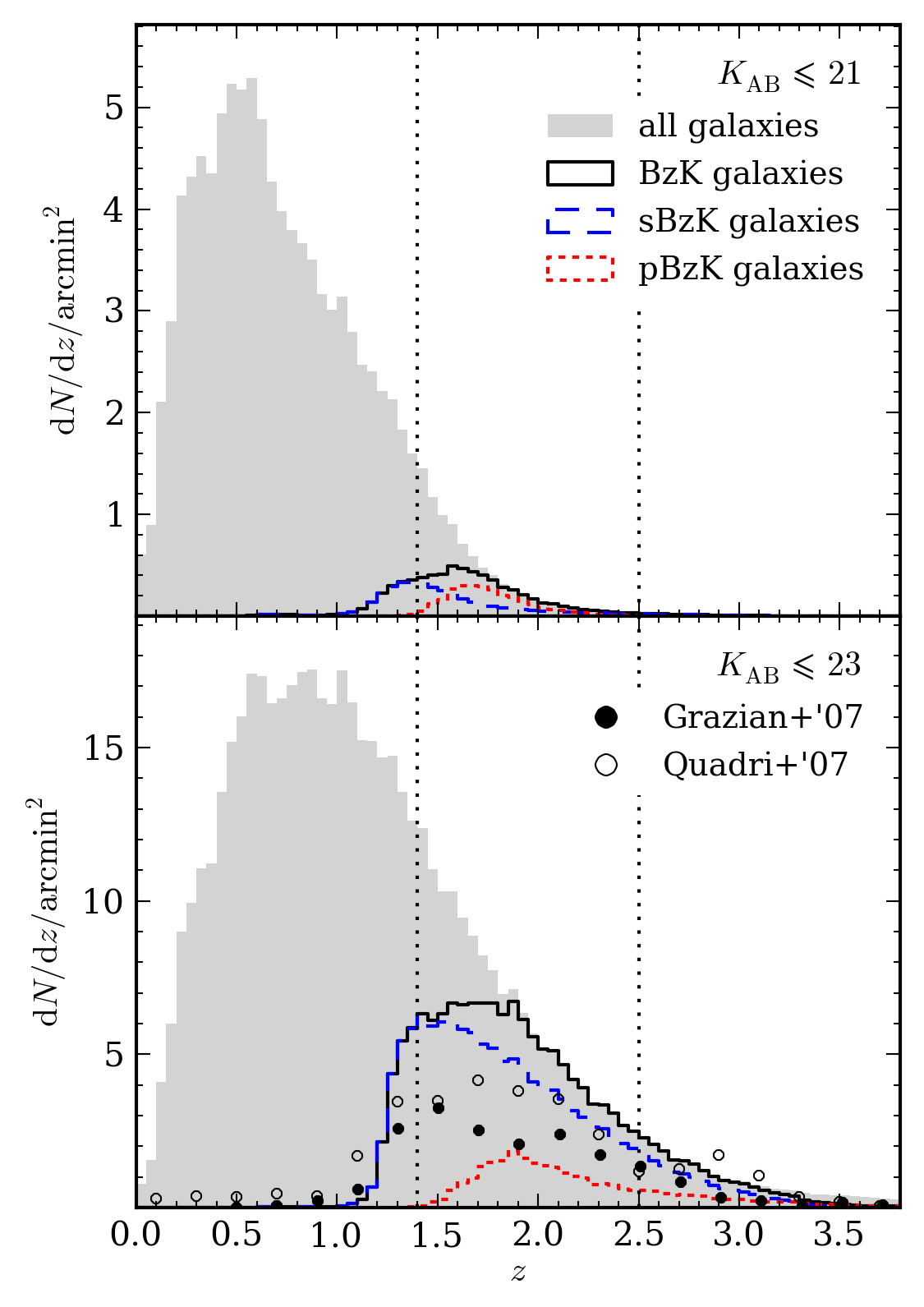}
\caption{The predicted redshift distributions of BzK selected galaxies
in the mock catalogue (black solid line) for two different ${\rm
K}$-band flux limits: ${\rm K_{AB}}\leqslant 21$ (top) and ${\rm
K_{AB}}\leqslant 23$ (bottom). For comparison, the redshift
distribution of all ${\rm K}$-band selected mock galaxies down to
these limits is shown by the grey shaded region. The limits of the
redshift range which the BzK technique was designed to probe are
indicated by the vertical dotted lines. Blue dashed and red dotted
histograms show the redshift distributions of sBzK and pBzK galaxies
respectively. In the bottom panel are plotted observed redshift
distributions for BzK galaxy samples with ${\rm K_{AB}}\leqslant 23.8$
and ${\rm K_{AB}}\leqslant 22.9$ from \protect\cite{Grazian07} and
\protect\cite{Quadri07} respectively.}
\label{fig:BzK_N(z)}
\end{center}
\end{figure}

In Fig.~\ref{fig:BzK_N(z)} we show the predicted redshift
distributions for BzK galaxies in the \GALFORM{} mock catalogue for
two example ${\rm K}$-band flux limits, ${\rm K_{AB}}\leqslant 21$ and
$23$. For comparison, we also show the redshift distribution for all
model galaxies brighter than the stated ${\rm K}$-band limit, and use
vertical dotted lines to indicate the redshift range which the BzK
technique is designed to probe, $1.4<z<2.5$.

It is clear from Fig.~\ref{fig:BzK_N(z)} that BzK galaxies probe the
high redshift tail of the redshift distribution of ${\rm K}$-selected
galaxies. For example, the predicted median redshift of the ${\rm
K_{\AB}}\leqslant 23$ sample is $z_{{\rm med}}\sim 1.2$, while the BzK
subsample has a higher median redshift of $z_{{\rm med}}\sim
1.9$. Moreover $\sim98$ per cent of the \GALFORM{} galaxies within
$2.0 \lesssim z <2.5$ are selected as BzK galaxies. However, in the
redshift range $1.4 < z\lesssim 2$, the fraction of galaxies selected
by the BzK technique decreases with decreasing redshift. For ${\rm
K_{AB}}\leqslant 21$ the fraction of galaxies at $z=1.4$ that are
recovered is $\sim20-25$ per cent, compared to $\sim 50$ per cent for
${\rm K_{AB}}\leqslant 23$. For ${\rm K_{AB}}\leqslant 21$ the
fraction of galaxies recovered reaches $50$ per cent at $z\sim1.55$.

\begin{figure*}
\begin{center}
\includegraphics[width=\textwidth]{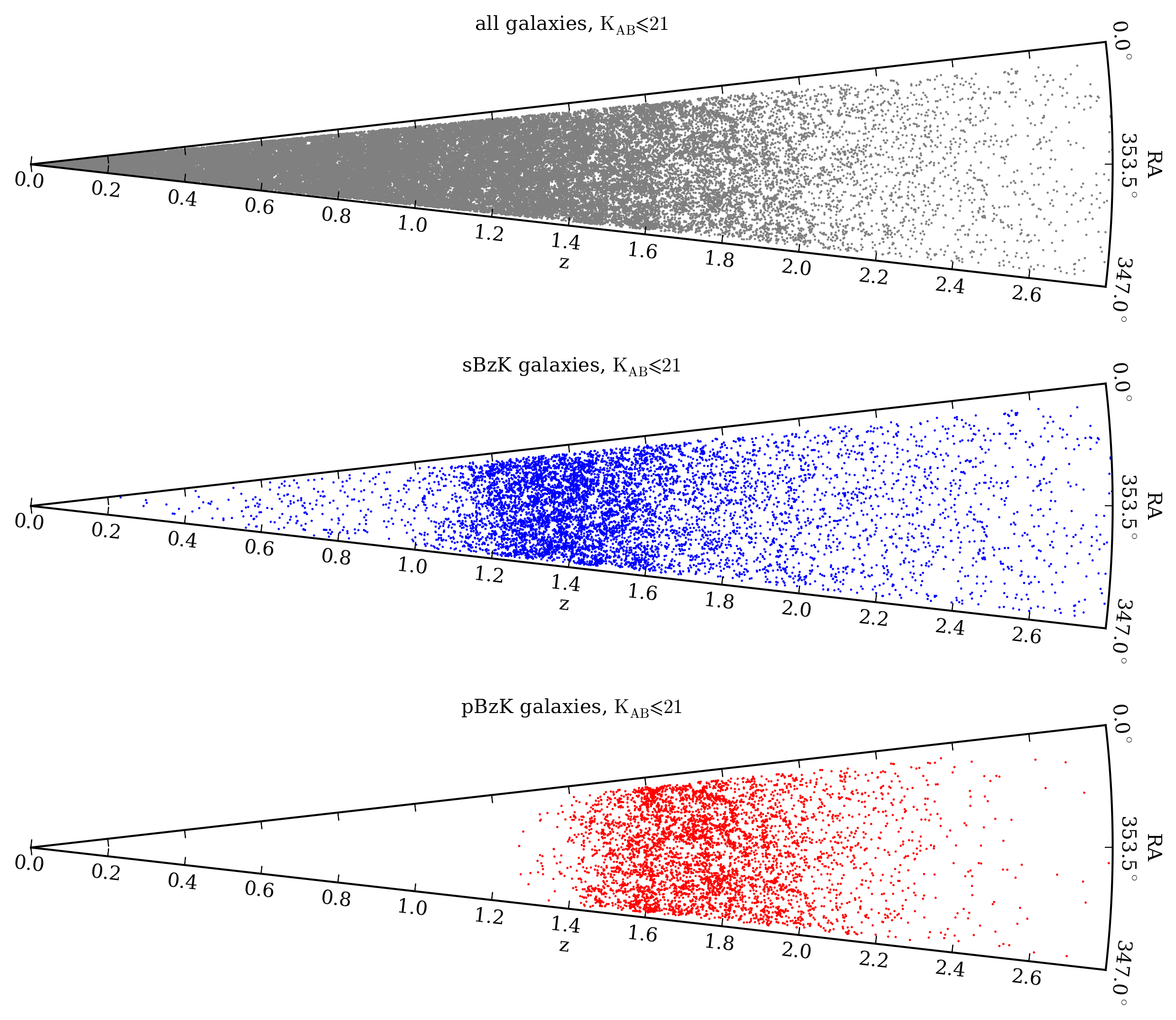}
\caption{Wedge plots showing a slice in redshift and right ascension,
$1^{\circ}$ wide in declination, of the predicted distribution of all
galaxies with ${\rm K_{\AB}}\leqslant 21$ (top) and the subsamples of
sBzKs (middle) and pBzKs (bottom).}
\label{fig:BzK_wedge}
\end{center}
\end{figure*}

In the lower panel of Fig.~\ref{fig:BzK_N(z)} we compare the redshift
distribution for our BzK selected galaxy sample with observed
photometric redshift distributions from \cite{Grazian07} and
\cite{Quadri07}, selected with ${\rm K_{AB}}\leqslant 23.8$ and ${\rm
K_{AB}}\leqslant 22.9$ respectively. The predicted BzK redshift
distribution has a median redshift, $z_{{\rm med}}\simeq 1.8$, that is
consistent with the median redshifts of the observed distributions,
$1.7\lesssim z_{{\rm med}}\lesssim 1.9$. As we have seen in the
left-hand panel of Fig.~\ref{fig:BzK_numbercounts}, the \GALFORM{}
model over-predicts the number counts of BzK galaxies at ${\rm
K_{\AB}}= 23$ and so, understandably, for all redshift bins within
$1.4<z<2.5$, the mock catalogue predicts a greater number of BzK
galaxies than is observed.

We can see from Fig.~\ref{fig:BzK_N(z)} that the redshift distribution
of sBzK galaxies consistently peaks at lower redshifts than the pBzK
distribution. This can also be seen in Fig.~\ref{fig:BzK_wedge}, which
shows the predicted large-scale distribution of ${\rm
K_{\AB}}\leqslant 21$ predicted galaxies and the subsamples of sBzK
and pBzK galaxies. Fig.~\ref{fig:BzK_wedge} shows that, while sBzK
galaxies can be selected at redshifts down to $z\sim 0$, pBzK galaxies
only start to appear at $z\sim 1.4$. In Fig.~\ref{fig:BzK_wedge} we
can also see that at $z\sim 2$ the pBzK galaxies appear to trace
filamentary structures compared to the sBzK galaxies, which appear to
be less clustered. Only for fainter limits of ${\rm K_{\AB}}\lesssim
23$, do sBzK galaxies begin to trace the filamentary structure at
$z\sim 2$. This suggests that the predicted spatial clustering of pBzK
galaxies is stronger than that for sBzKs, in agreement with
observations \citep[e.g. ][]{Kong06,Hartley08}.

\begin{figure*}
\begin{center}
\includegraphics[width=17.2cm]{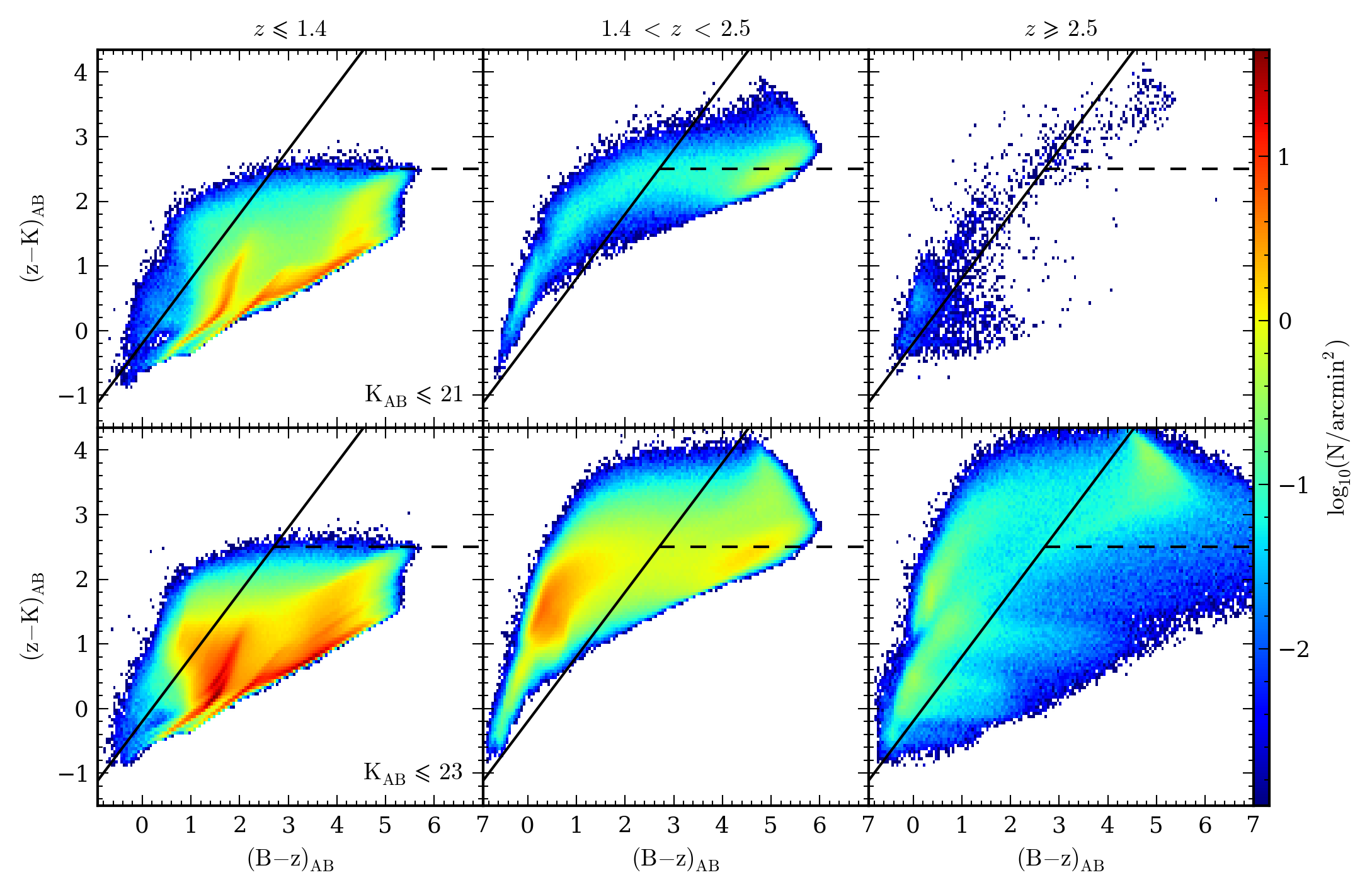}
\caption{The distribution of synthetic galaxies in the BzK colour
plane for two $K$-band flux limits, ${\rm K_{\AB}}\leqslant 21$ (top
row) and ${\rm K_{\AB}}\leqslant 23$ (bottom row). The columns
correspond to three different redshift ranges: ${\rm z}\leqslant1.4$
(left), $1.4<{\rm z}<2.5$ (the redshift interval which the BzK
technique was designed to select, middle) and ${\rm z}\geqslant2.5$
(right). The black solid line and dashed line correspond to the sBzK
and pBzK cuts of \protect\cite{Daddi04b} respectively. The colour
shading indicates the surface density of galaxies on the mock sky, as
shown by the scale on the right-hand side.}
\label{fig:BzK_planes}
\end{center}
\end{figure*}

\subsection{Efficiency of the BzK selection}
\label{sec:efficiency_of_BzK}

The BzK technique was designed to select galaxies within $1.4<z<2.5$
and to separate them into star-forming and passively evolving
subsamples. To assess the effectiveness with which the BzK technique
achieves these goals, we consider the completeness
($\S$~\ref{sec:K_completeness}) and contamination
($\S$~\ref{sec:contamination}) of a BzK galaxy sample selected from
the \GALFORM{} mock catalogue.

\subsubsection{$K$-band completeness}
\label{sec:K_completeness}

In this section we explore the fraction of ${\rm K}$-band selected
galaxies at $1.4<z<2.5$ that are actually picked up when using the BzK
selection technique for galaxies in the \GALFORM{} mock catalogue. For
this purpose we compare the predicted number counts of BzK galaxies,
presented in the left-hand panel of Fig.~\ref{fig:BzK_numbercounts},
with the total number counts of ${\rm K_{\AB}}$-band selected galaxies
that lie within the target redshift range, shown by the dotted line in
the same panel of Fig.~\ref{fig:BzK_numbercounts}. Faintwards of ${\rm
K_{AB}}\sim 19.5$ the predicted BzK counts are in good agreement with
the counts of $1.4<z<2.5$ galaxies, indicating that the BzK selection
is an effective probe of the galaxy population at this
epoch. 

In Fig.~\ref{fig:BzK_planes} we show the ${\rm (B-z)}$ vs. ${\rm
(z-K)}$ plane populated by \GALFORM{} galaxies within three different
redshift regimes, $z\leqslant 1.4$ (left column), $1.4<z<2.5$ (middle
column) and $z\geqslant 2.5$ (right column). The distribution is shown
for our two example ${\rm K}$-band flux limits: ${\rm
K_{\AB}}\leqslant 21$ (top row) and ${\rm K_{\AB}}\leqslant 23$
(bottom row). We define the completeness of the BzK technique as the
fraction of all galaxies in $1.4<z<2.5$ that lie in either of the BzK
regions in the ${\rm (B-z)}$ vs. ${\rm (z-K)}$ plane. About a quarter
of the galaxies brighter than ${\rm K_{\AB}}=21$ within $1.4<z<2.5$
lie outside of the BzK regions. The distribution has two clear peaks,
one at $({\rm B-z})\sim 0$, which we will refer to as the star-forming
peak, and the other at $({\rm B-z})\sim 5$, which we will refer to as
the passively evolving peak. The star-forming peak falls well within
the sBzK region, while the passively evolving peak lies just outside
the pBzK region. This would explain the under-prediction of the pBzK
number counts for ${\rm K_{\AB}}\lesssim 20$. However, for ${\rm
K_{\AB}}\leqslant 23$, the star-forming peak dominates the galaxy
population suggesting that for fainter ${\rm K}$-band limits the BzK
selection provides a more complete sample of the $1.4<z<2.5$ galaxy
population.

\begin{figure}
\begin{center}
\includegraphics[width=8.6cm]{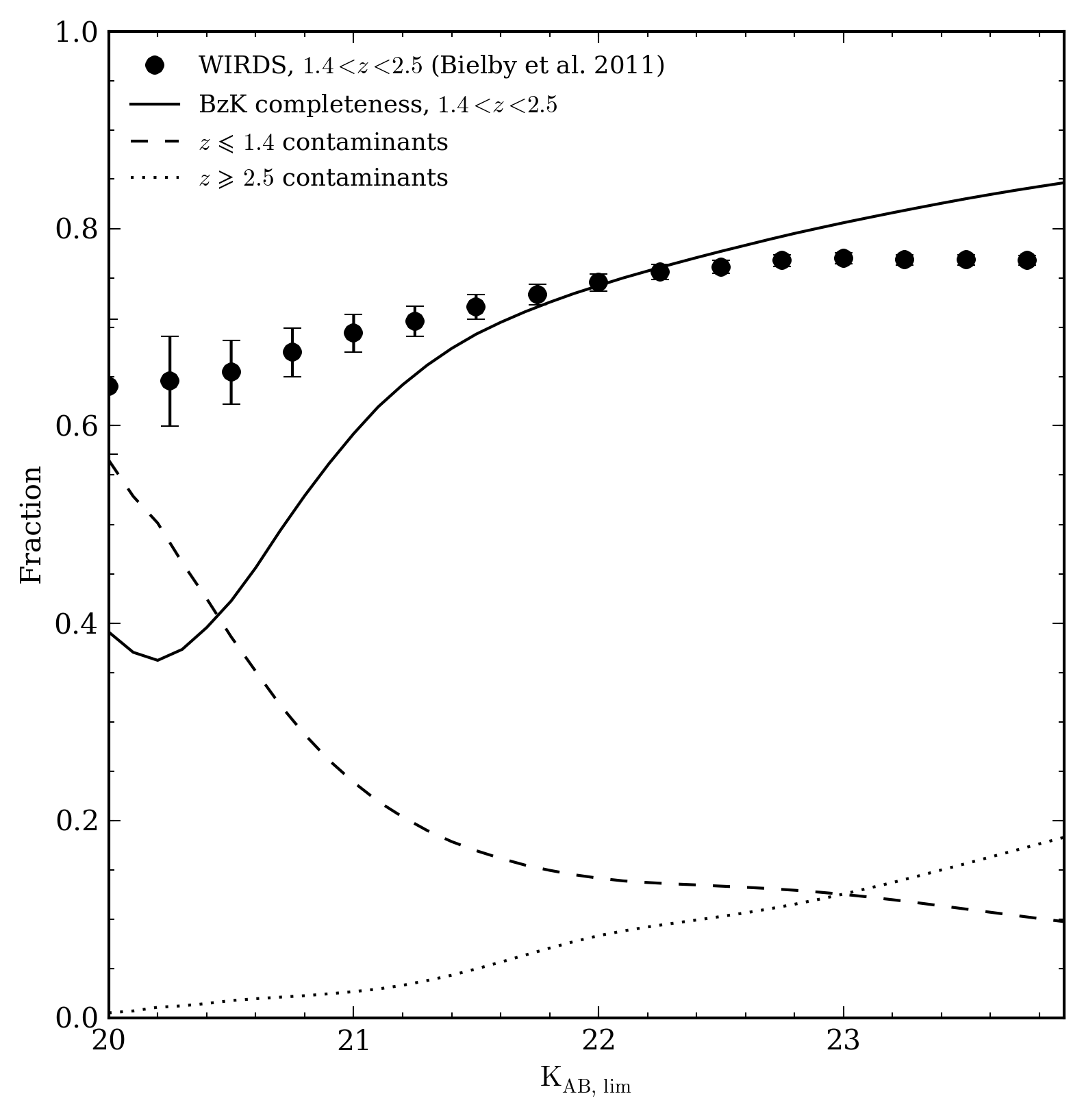}
\caption{The efficiency of the BzK selection as a function of $K$-band
limiting magnitude, ${\rm K_{\AB ,lim}}$. The solid line shows the
predicted fraction of \GALFORM{} galaxies within $1.4< z< 2.5$, with
${\rm K_{\AB}} \leqslant {\rm K_{\AB ,lim}}$, that are identified as
BzK galaxies. The filled circles correspond to completeness estimates
for observed galaxies in the WIRCam Deep Survey \protect\citep[WIRDS,
][]{Bielby11*} that have been calculated in the same way as the
\GALFORM{} predictions. The error bars shown correspond to Poisson
errors. The dashed and dotted lines show the predicted fraction of
interlopers at $z\leqslant 1.4$ and $z\geqslant 2.5$ respectively, as
a function of ${\rm K}$-band limiting magnitude.}
\label{fig:BzKcompleteness}
\end{center}
\end{figure}

The completeness of the BzK technique, as a function of the limiting
${\rm K}$-band magnitude of the galaxy sample, is shown in
Fig.~\ref{fig:BzKcompleteness}. Here, the data points show the BzK
completeness estimates from \cite{Bielby11*}, who applied the BzK
selection to an input catalogue of $\sim1.8$ million ${\rm K}$-band
galaxies, with photometric redshifts ($\sigma_{{\rm
\Delta}z/(1+z)}\lesssim 0.03$), from the WIRCam Deep Survey (WIRDS).
We can clearly see in Fig.~\ref{fig:BzKcompleteness} that the BzK
completeness increases with fainter ${\rm K}$-band limiting
magnitude. The same trend is seen for the completeness predictions for
\GALFORM{} galaxies, shown by the solid line, with $\sim55$, $\sim73$
and $\sim80$ per cent of $1.4<z<2.5$ galaxies being recovered for
${\rm K_{\AB}}\leqslant21$, $22$ and $23$ respectively. Therefore, for
faint ${\rm K}$-band limits (${\rm K_{AB,lim}}\gtrsim 22$), the BzK
technique is consistently selecting $75$ to $80$ per cent of the
galaxies within $1.4<z<2.5$. However, for a very bright limit of ${\rm
K_{\AB}}\leqslant 20$ the technique identifies less than half of the
galaxy population within the target redshift range. For $21 \lesssim
{\rm K_{AB,lim}}\lesssim 22$ the completeness estimates from the
\GALFORM{} mock catalogue are in very good agreement with the WIRDS
estimates.

\subsubsection{Contamination}
\label{sec:contamination} 
We now explore the predicted numbers of galaxies outside the redshift
range, $1.4<z<2.5$, that are picked up by the BzK selection. As we can
see from Fig.~\ref{fig:BzK_planes}, it is not possible to ever have a
sample of BzK selected galaxies that is entirely free of contamination
from interlopers with low, $z \leqslant 1.4$, or high redshift, $z
\geqslant 2.5$, which are classified as BzK galaxies. The left-hand
column of Fig.~\ref{fig:BzK_planes} shows that low redshift
interlopers are typically classified as sBzK galaxies, while the ${\rm
(z-K)}$ cut used in Eq.~\ref{eq:pBzK} successfully eliminates low
redshift pBzK galaxies. High redshift interlopers, shown in the
right-hand column of Fig.~\ref{fig:BzK_planes}, appear to be more
evenly distributed between the sBzK and pBzK regions and thus more
difficult to remove. 

The fractions of low and high redshift interlopers, as a function of
${\rm K}$-band limiting magnitude, are shown in
Fig.~\ref{fig:BzKcompleteness} by the dashed and dotted lines
respectively. From Fig.~\ref{fig:BzKcompleteness} we can see that by
applying a bright ${\rm K_{AB}}\leqslant 20$ selection to our
\GALFORM{} mock catalogue, the BzK technique selects approximately
equal numbers of galaxies with $1.4<z<2.5$ and $z\leqslant
1.4$. Pushing the ${\rm K}$-band selection limit to fainter magnitudes
leads to a decrease in the fraction of low redshift contamination as
an increasing number of galaxies within $1.4<z<2.5$ become visible at
fainter ${\rm K}$-band limits. Fig.~\ref{fig:BzK_N(z)} shows clearly
how the redshift distribution of BzK galaxies develops a sharper low
redshift cut-off as the flux limit is made fainter. By ${\rm
K_{AB,lim}}\sim 21.5$, the low redshift contamination has fallen to
$\sim18$ per cent. For fainter flux limits the low redshift
contamination decreases much more slowly, reaching $\sim10$ per cent
by ${\rm K_{AB,lim}}\sim 24$.

As expected, the fraction of high redshift interlopers increases
steadily with increasingly faint limiting ${\rm K}$-band magnitude,
though it stays well below $\sim 20$ per cent. By ${\rm
K_{AB,lim}}\sim 23.2$, the contribution from low redshift and high
redshift contamination is approximately equal at $\sim12$ per cent,
with high redshift interlopers dominating the contamination at fainter
limiting magnitudes.

Although we have not included the effect of the inter-galactic medium
(IGM) attenuation in this particular lightcone, we have checked its
effect on the galaxy $({\rm B}-{\rm z})$ and $({\rm z}-{\rm K})$
colours. We have examined the $({\rm B}-{\rm z})$ vs. $({\rm z}-{\rm
  K})$ plane at discrete redshift snapshots: $z\sim2.0$, $z\sim2.5$
and $z\sim3$. We find that the IGM attenuation has only a modest
affect on the number of BzK galaxies at $z\gtrsim 3$, which is well
into the high redshift tail of the galaxy redshift distribution.

\subsubsection{Dependence on $B$-band depth}
\label{sec:Bband_depth}

As we have seen in $\S$\ref{sec:predicted_numbers}, there is evidence
that the ability of the BzK technique to distinguish between
star-forming and passive galaxies within $1.4<z< 2.5$ is dependent
upon the $B$-band depth of the galaxy sample. For example,
\cite{Grazian07} determined that $22$ per cent of their sample of sBzK
galaxies had SEDs typical of passive galaxies rather than star-forming
galaxies. A significant number of these galaxies were undetected in
the $B$-band and had their $(B-z)$ colours estimated using a $1\sigma$
$B$-band upper limit, which resulted in their $(B-z)$ colours being
too blue. \citeauthor{Grazian07} concluded that, for faint $K$-band
selected galaxies with very red $(z-K)$ colours, a lack of deep
$B$-band photometry will lead to many pBzK galaxies being incorrectly
classified as sBzK galaxies.

\begin{figure}
\begin{center}
\includegraphics[width=8.6cm]{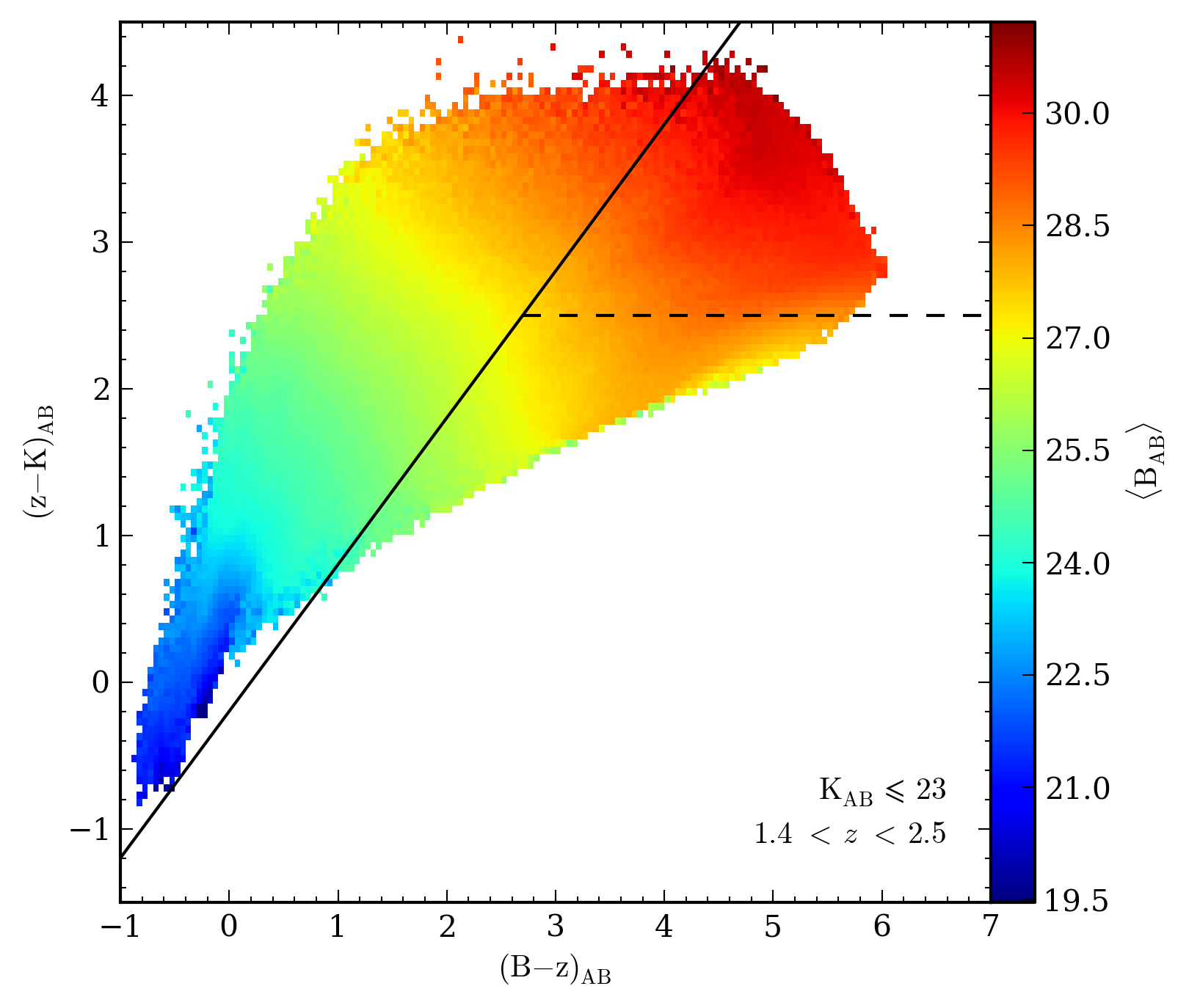}
\caption{The variation in the median $B$-band apparent magnitude of
galaxies with position in the ${\rm (B-z)}$ vs. ${\rm (z-K)}$ plane.
The distribution shown corresponds to \GALFORM{} galaxies, within
$1.4<z<2.5$, selected to have ${\rm K_{\AB}}\leqslant23$ and placed
into 2-dimensional bins spanning the BzK colour-colour space. The bins
are coloured according to the median $B$-band magnitude of the
galaxies in that bin, as shown by the colourbar.}
\label{fig:Bband_limits}
\end{center}
\end{figure}

In Fig.~\ref{fig:Bband_limits}, we show the variation of the median
${\rm B}$-band apparent magnitude with position in the ${\rm (B-z)}$
vs. ${\rm (z-K)}$ plane for ${\rm K_{\AB}}\leqslant 23$ galaxies in
the \GALFORM{} mock catalogue. The trend towards fainter ${\rm
B}$-band magnitudes for redder $({\rm B-z})$ and $({\rm z-K})$ colours
is immediately clear and supports the need for deep ${\rm B}$-band
photometry to probe the faint pBzK population.

We apply a ${\rm B}$-band detection limit of ${\rm
B_{AB,lim}}\leqslant 26$ to a ${\rm K_{\AB}}\leqslant 23$ sample of
galaxies, by assuming that galaxies with ${\rm B}$-band magnitudes
fainter than ${\rm B_{AB,lim}}$ are undetected and so have ${\rm
B_{AB}} = {\rm B_{AB,lim}}$. By doing this, we find that only
$\sim0.3$ per cent of galaxies within $1.4<z<2.5$ are classified as
pBzK galaxies. Making the ${\rm B}$-band limit fainter leads to a
larger fraction of pBzK galaxies. With upper limits of ${\rm
B_{AB,lim}}\leqslant 27$, $28$ and $29$ we find that $\sim3$, $\sim9$
and $\sim15$ per cent of $1.4<z<2.5$ galaxies respectively are
classified as pBzK galaxies. An upper limit of ${\rm
B_{AB,lim}}\leqslant 30$ leads to the same number of pBzK galaxies
being recovered ($\sim16$ per cent) as when applying the ${\rm
K_{AB}}\leqslant 23$ selection in isolation.

A bright ${\rm B}$-band limit will also lead to galaxies that should
not be classified as BzK galaxies being scattered into the sBzK region
of the ${\rm (B-z)}$ vs. ${\rm (z-K)}$ plane. As we have seen in
$\S$~\ref{sec:K_completeness}, for a ${\rm K_{AB}}\leqslant 23$
selected galaxy sample, the BzK technique selects $\sim80$ per cent of
$1.4<z<2.5$ galaxies. If we apply ${\rm B}$-band detection limits of
${\rm B_{AB,lim}}\leqslant 26$, $27$ and $28$ we find that the BzK
technique selects $\sim95$, $\sim87$ and $\sim80$ per cent of galaxies
within $1.4<z<2.5$ respectively. 

We conclude that adopting a fainter ${\rm B}$-band limit should
improve the ability of the BzK technique to distinguish between
star-forming and passively evolving galaxies.

\subsection{The predicted properties of BzK galaxies}
\label{sec:BzK_properties}

\subsubsection{Stellar mass}
\label{sec:BzK_stellar_mass}
From the upper left panel of Fig.~\ref{fig:BzK_property_planes} we can
see that the predicted stellar masses of galaxies in a ${\rm
K_{\AB}}\leqslant 23$ selected BzK sample range from $\sim 10^9
h^{-1}\Msol$ to $\sim 10^{11} h^{-1}\Msol$, with the more massive
galaxies typically having redder $({\rm z-K})$ colours.

\begin{figure}
\begin{center}
\includegraphics[width=8.6cm]{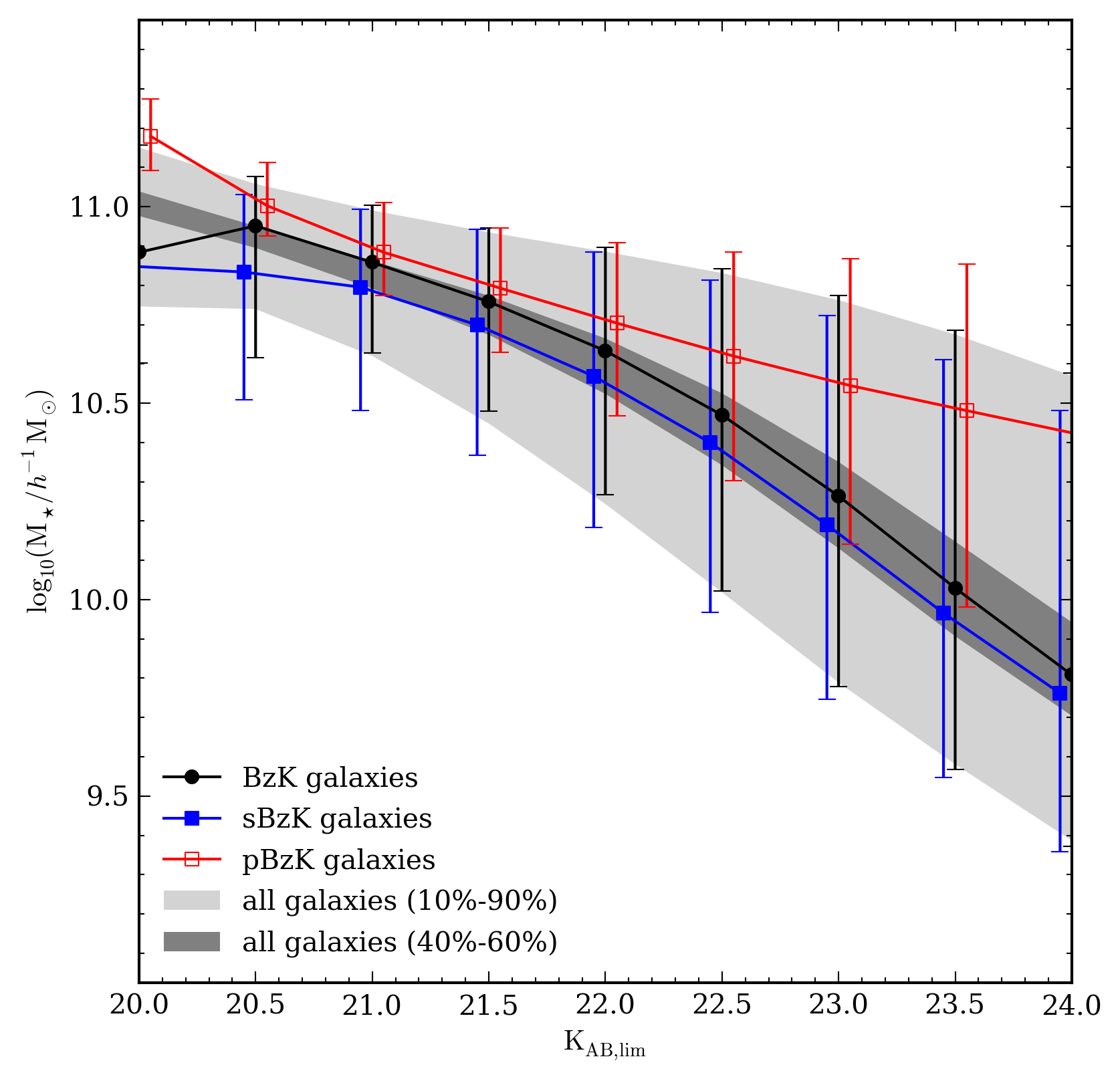}
\caption{The predicted stellar mass of galaxies with redshift
$1.4<z<2.5$, as a function of ${\rm K}$-band limiting magnitude for
all BzK galaxies (black circles), sBzK galaxies (blue, filled squares)
and pBzK galaxies (red, open squares). Data points correspond to
median values and error bars show the $10$ and $90$ percentiles. For
clarity, the data points for the sBzK and pBzK values have been offset
horizontally. The light and dark grey regions show the $10-90$ and
$40-60$ percentiles for all galaxies brighter than the ${\rm K}$-band
flux limit (i.e. irrespective of whether they are BzK selected).}
\label{fig:stellar_mass_medians}
\end{center}
\end{figure}

We show in Fig.~\ref{fig:stellar_mass_medians} the distribution of
stellar masses for all ${\rm K}$-band selected galaxies (within
$1.4<z<2.5$). The median stellar mass of BzK selected galaxies is in
excellent agreement with the distribution for all ${\rm K}$-band
selected galaxies for all flux limits fainter than ${\rm
K_{AB,lim}}\sim 21$. Additionally, the $10$ and $90$ percentiles of
the BzK distribution consistently match the $10$ and $90$ percentiles
for the stellar mass distribution of the whole galaxy population.

Early studies of BzK galaxies, using ${\rm K}$-band limits of ${\rm
K_{AB}}\lesssim 22$, inferred BzK galaxies to be very massive, with
typical stellar masses\footnote{The quoted value for the observed mass
has been multiplied by a factor of 1.4 \citep{Fontana04} in order to
account for the change from \cite{Salpeter55} IMF, used in
observational studies, to the \cite{Kennicutt83} IMF used for the
study presented here.}  of $M_{\star}\gtrsim 5\times
10^{10}h^{-1}\Msol$
\citep{Daddi04a,Daddi04b,Daddi05a,Daddi05b,Reddy05,Kong06,Blanc08}. In
Fig.~\ref{fig:stellar_mass_medians} we show the median stellar mass of
BzK selected galaxies (within $1.4<z<2.5$) as a function of the ${\rm
K}$-band flux limit. For ${\rm K_{\AB}}\leqslant 22$, the distribution
of BzK stellar masses in the mock catalogue is consistent with
observations. Increasing the depth in the ${\rm K}$-band, leads to a
shift in the distribution towards smaller stellar masses, with a
median BzK stellar mass of $\sim 10^{10}h^{-1}\Msol$ being reached at
${\rm K_{AB,lim}}\sim 23.5$.

We also show in Fig.~\ref{fig:stellar_mass_medians}, the breakdown of
the distribution into sBzK and pBzK galaxies. It is immediately clear
that, typically, pBzK galaxies are more massive than sBzK galaxies,
with the difference between the medians increasing towards fainter
${\rm K}$-band limits.

We conclude that BzK selected galaxies appear to provide a
representative sample of the galaxy stellar masses at $1.4<z<2.5$ and
do not appear to be significantly biased towards either very high or
low mass galaxies.

\subsubsection{Star Formation Rate}
\label{sec:BzK_sfr}

As we have already seen in Fig.~\ref{fig:BzK_property_planes}, there
is clear trend in the predicted SFR of galaxies across the ${\rm
  (B-z)}$ vs. ${\rm (z-K)}$ plane. In the extremes of the distribution
we find that many sBzK selected galaxies are predicted to have SFRs of
$\sim 100\,h^{-1}\Msolyr$ or more, while many pBzK selected galaxies
have SFRs of effectively zero. We find that trend in the specific
star-formation rate (sSFR, equal to the star-formation rate of a
galaxy divided by its stellar mass) across the ${\rm (B-z)}$ vs. ${\rm
  (z-K)}$ plane is almost identical to that of the star-formation
rate.

\begin{figure}
\begin{center}
\includegraphics[width=8.6cm]{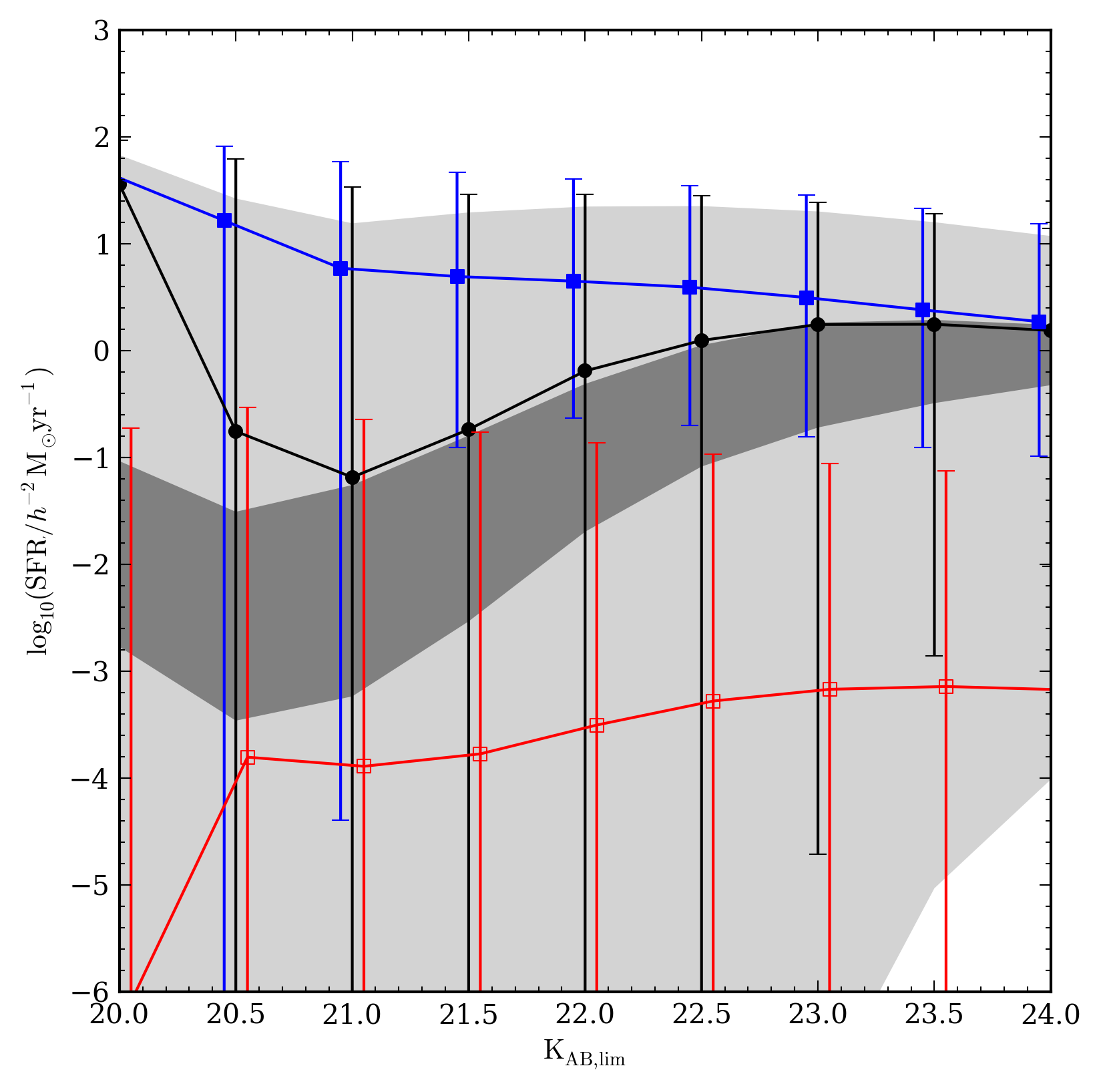}
\caption{The predicted star-formation rate as a function of ${\rm
K}$-band limiting magnitude for galaxies in $1.4<z<2.5$. The symbols
and shaded regions are the same as in
Fig.~\ref{fig:stellar_mass_medians}.}
\label{fig:sfr_medians}
\end{center}
\end{figure}

In Fig.~\ref{fig:sfr_medians} we show the distribution of SFRs for
BzK, sBzK and pBzK galaxies, as well as for all ${\rm K}$-band
selected galaxies, as a function of ${\rm K}$-band flux limit, in the
redshift range $1.4<z<2.5$.

For ${\rm K_{\AB}}\gtrsim 21$, the median SFR for BzK galaxies is in
reasonable agreement with the distribution for the whole galaxy
population, though is perhaps slightly biased towards higher
SFRs. This would, at first, suggest that the BzK selection is missing
a fraction of the passive galaxy population, particularly since we
have shown in Fig.~\ref{fig:BzK_numbercounts} that the \GALFORM{} mock
catalogue matches the number of sBzK galaxies but under-predicts the
number of bright pBzK galaxies. It is possible that some fraction of
these faint pBzK galaxies are dusty star-forming galaxies that have
been mis-classified as being passive. For ${\rm K_{\AB}}\leqslant 21$,
we find that $\sim20$ per cent of the pBzK selected galaxies in the
\GALFORM{} mock catalogue have SFRs $\gtrsim 0.1\,
h^{-1}\Msolyr$. Interestingly, the typical SFR of pBzK galaxies
remains approximately constant, at $\sim 10^{-4}-10^{-3}\,
h^{-1}\Msolyr$, with increasing ${\rm K}$-band depth (though the
distribution is very broad). For ${\rm K_{AB,lim}}\gtrsim 23$, the
typical SFR of sBzK galaxies also appears to remain almost constant at
$\sim 1-10\, h^{-1}\Msolyr$.

In Fig.~\ref{fig:sfr_medians}, we see that the model predicted median
SFR of BzK galaxies with ${\rm K_{\AB}}\leqslant 22$ to be $\sim
1\,h^{-1}\Msolyr$. However, observational studies of BzK galaxies with
${\rm K_{\AB}}\lesssim 22$ concluded many of these bright BzK galaxies
to be starbursting galaxies, with SFRs of $\sim 50\,h^{-2}\Msolyr -
100\,h^{-2}\Msolyr$ \citep[e.g. ][]{Daddi04b,Kong06,Blanc08}. A
possible explanation for this discrepancy is the overly efficient shut
down of gas cooling by AGN feedback in the \cite{Bower06} model, which
has been previously suggested by \cite{Gonzalez-Perez09}.

Based upon the model predictions however, we predict that, towards
fainter ${\rm K}$-band limiting magnitudes, the BzK technique is
typically selecting galaxies with SFRs that are consistent with the
median SFRs of the galaxy population within $1.4<z<2.5$.

We note that a trend similar to that seen in the predicted median SFR
is seen in the median values of the sSFR of BzK selected galaxies. At
faint ${\rm K}$-band limits, the median sSFR of BzK and sBzK galaxies
tends towards $\sim10^{-10}\,{\rm yr}^{-1}$. The \GALFORM{} model
predicts this value to be typical for ${\rm K}$-band selected galaxies
at $1.4<z<2.5$. As with the median SFR, the median sSFR tends towards
a constant value of $\sim10^{-14}\,{\rm yr}^{-1}$.

\subsubsection{Metallicity}
\label{sec:metallicity}

We have already seen in $\S$~\ref{sec:BzK_intro} that the metallicity
of ${\rm K}$-band selected galaxies varies with position in the $({\rm
B-z})$ vs. $({\rm z-K})$ plane. From the lower left-hand panel of
Fig.~\ref{fig:BzK_property_planes} we can see that galaxies with the
reddest $({\rm z-K})$ colours (typically pBzK and faint sBzK galaxies)
are in general the most metal rich.

\begin{figure}
\begin{center}
\includegraphics[width=8.6cm]{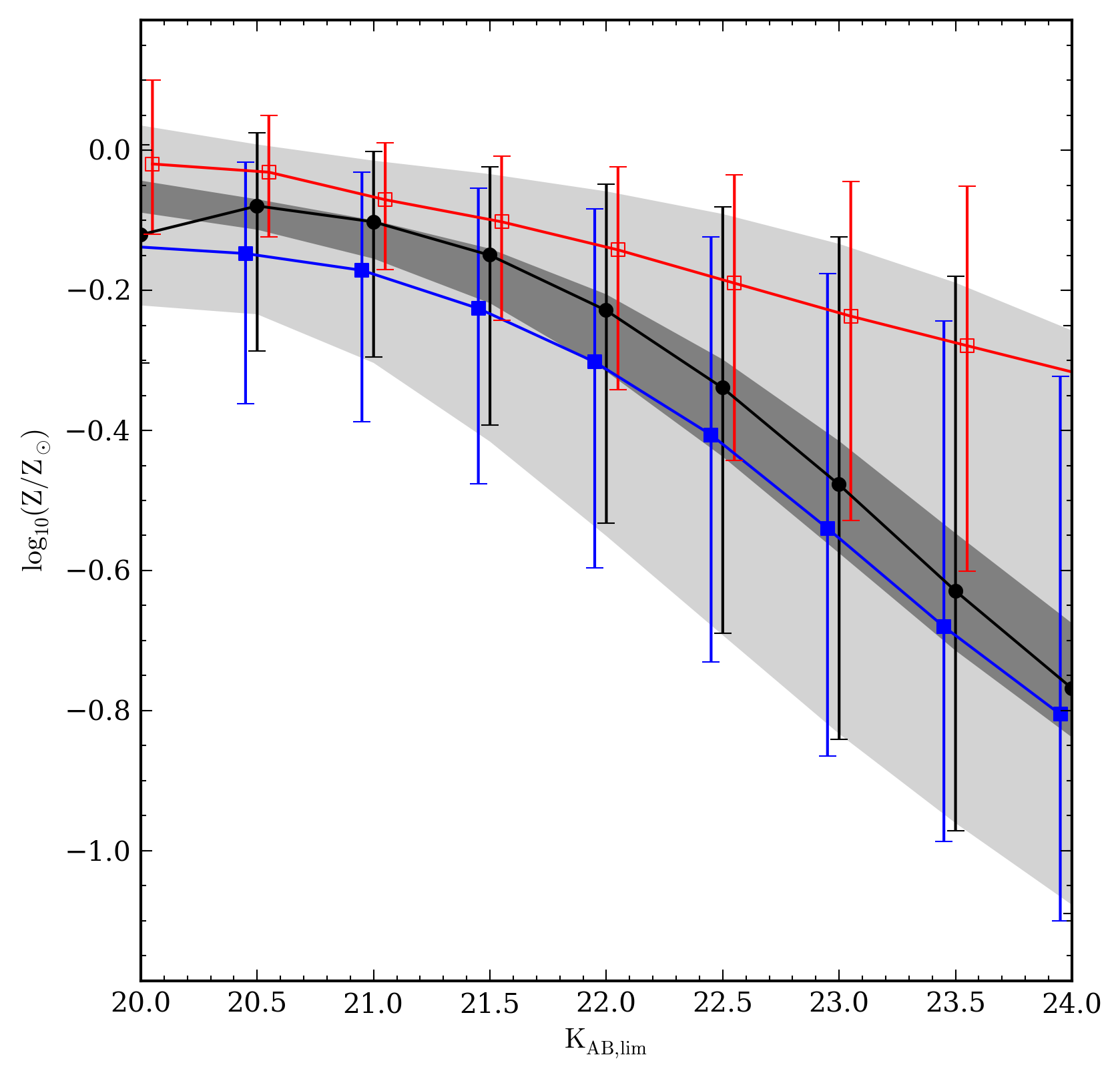}
\caption{The predicted stellar metallicity as a function of ${\rm
K}$-band limiting magnitude for galaxies in $1.4<z<2.5$. The symbols
and shaded regions are the same as in
Fig.~\ref{fig:stellar_mass_medians}.}
\label{fig:metal_medians}
\end{center}
\end{figure}

In Fig.~\ref{fig:metal_medians} we show the metallicity distribution
for BzK selected galaxies within $1.4<z<2.5$. For all ${\rm K}$-band
limits considered the metallicity distribution of BzK selected
galaxies is in good agreement with the metallicity distribution for
all ${\rm K}$-band selected galaxies. The trend in the metallicity
distribution as a function of ${\rm K}$-band flux limit is very
similar to the trend seen in the stellar mass distribution in
Fig.~\ref{fig:stellar_mass_medians}. For brighter ${\rm K}$-band flux
limits, one would predict to recover BzK galaxies with higher
metallicities. For the brightest flux limit considered, the median
metallicities for BzK galaxies falls below that for all galaxies. As
with the stellar mass distribution, this is due to \GALFORM{}
under-predicting the counts of bright pBzK galaxies, which one would
expect to be metal-rich. The distributions for the separate sBzK and
pBzK subsets show that for any ${\rm K}$-band depth, pBzK galaxies
will typically be more metal-rich than sBzK galaxies, though the
distributions for the two subsets do overlap.

\subsubsection{Age}
\label{sec:age}

In the lower right-hand panel of Fig.~\ref{fig:BzK_property_planes} we
show the median stellar mass weighted age of galaxies in the $({\rm
B-z})$ vs. $({\rm z-K})$ plane. We find that the oldest galaxies
occupy the region where the density of passive galaxies peaks just
below the pBzK region, as can be seen in the middle column of
Fig.~\ref{fig:BzK_planes}. The vast majority of the oldest galaxies,
with ages above $2\,{\rm Gyr}$, that fall outside the pBzK region lie
within the redshift interval $1.4<z<2.5$. This is due to the finite
width of the $4000{\rm \AA}$ break, which at $z\sim1.4$ is beginning
to enter the response curve of the ${\rm z}$-band, thus making the
$({\rm z-K})$ colours of these galaxies bluer. Above $z=2$, all of the
galaxies with ages above approximately $1.5 \, {\rm Gyr}$ lie well
within the pBzK region on the colour plane. We have checked that the
$({\rm z-K})$ colours of the galaxies are not significantly affected
by changing between a \cite{Kennicutt83} and a \cite{Salpeter55} IMF.

\begin{figure}
\begin{center}
\includegraphics[width=8.6cm]{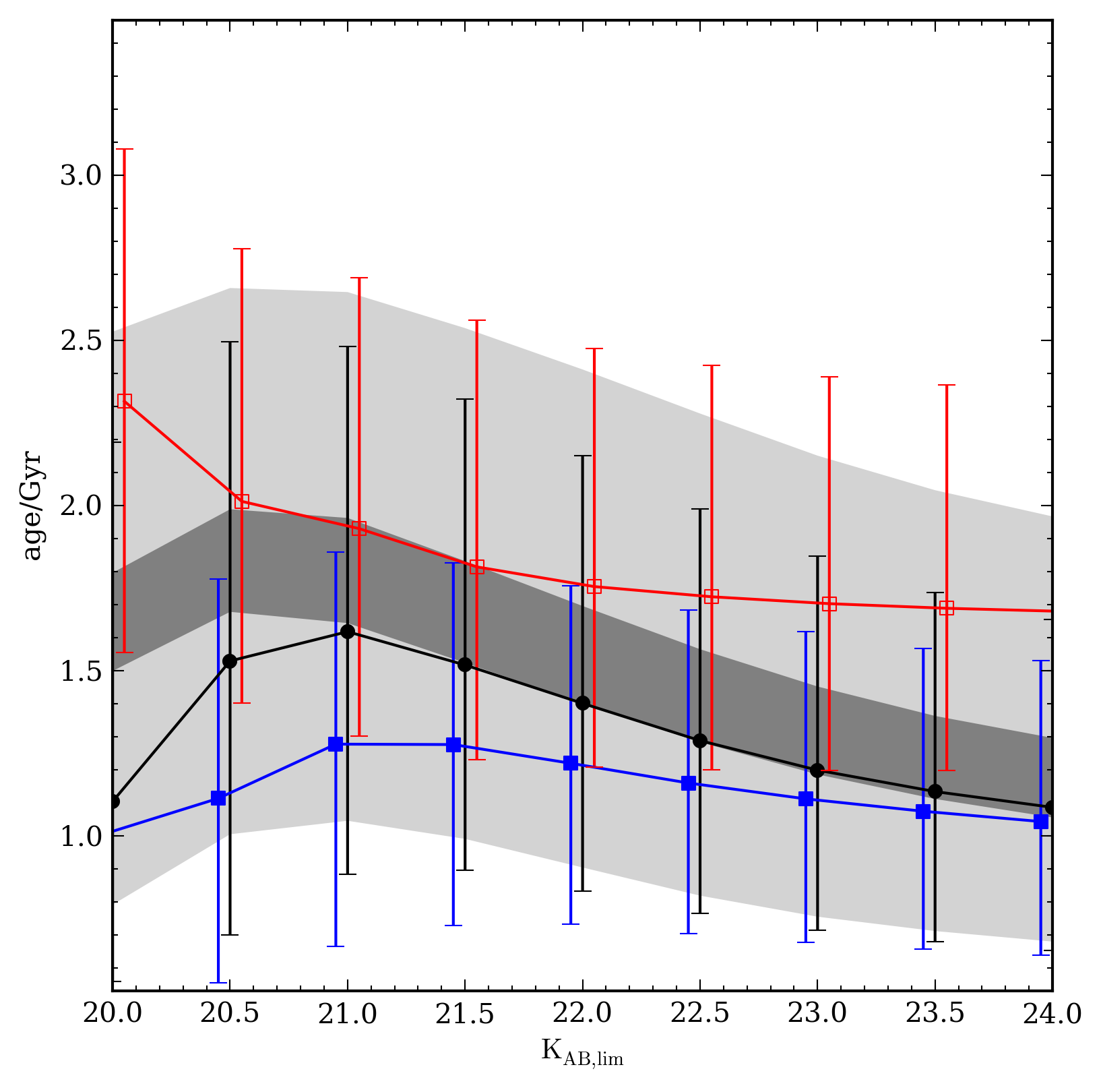}
\caption{The predicted stellar mass weighted age as a function of
${\rm K}$-band limiting magnitude for galaxies in $1.4<z<2.5$. The
symbols and shaded regions are the same as in
Fig.~\ref{fig:stellar_mass_medians}.}
\label{fig:age_medians}
\end{center}
\end{figure}

We show in Fig.~\ref{fig:age_medians} the distribution of the stellar
mass weighted ages for all ${\rm K}$-band selected galaxies and for
those that are BzK-selected. Like the distribution of SFRs, the
distribution of ages of BzK galaxies is in reasonable agreement with
the age distribution for all ${\rm K}$-band selected galaxies, though
appears to be slightly biased towards younger galaxies. As with the
SFR distribution, we see that the typical ages of sBzK and pBzK
galaxies remain approximately constant (at $\sim 1.1 {\rm Gyr}$ and
$\sim 1.7 {\rm Gyr}$ respectively) for ${\rm K_{AB,lim}}\gtrsim 22$.

\subsubsection{Dust}
\label{sec:dust}
Reddening due to dust can mimic a large break at $4000{\rm \AA}$ in
the spectra of star-forming $z\lesssim 1.4$ galaxies
\citep{Kriek06a,Kriek11}.  However, many authors have argued that the
effectiveness of the BzK colour selection is not significantly
affected by dust extinction \citep[e.g.]
[]{Daddi04b,Kong06,Hayashi07,Grazian07,Hartley08,Hayashi09,Lin11*}.

For the ${\rm K}$-band limits considered in Fig.~\ref{fig:BzK_planes}
we find that in the presence of dust the distribution of $1.4<z<2.5$
galaxies in the $({\rm B-z})$ vs. $({\rm z-K})$ plane remains
relatively unchanged, aside from an increased scatter in the sBzK
galaxy population towards redder $({\rm z-K})$ colours. We find that
for ${\rm K_{\AB}}\leqslant 23$ the presence of dust reddens the
colours of BzK galaxies, within $1.4<z<2.5$, by $\Delta({\rm
B-z})\sim0.15$ and $\Delta({\rm z-K})\sim0.3$. The presence of dust
appears to have a greater effect on the median colours of sBzK
galaxies, as we find a negligible change in the median colours of pBzK
galaxies. We can see this also in Fig.~\ref{fig:BzK_numbercounts}
where the number counts of sBzK galaxies without dust extinction are
boosted by $\sim 1$ dex, while the pBzK number counts remain the
same. The reduction in the sBzK counts when dust extinction is
included is likely due to dust reddening the ${\rm (B-z)}$ colours of
star-forming galaxies (with $({\rm z-K})_{\AB}<2.5$) and scattering
them out of the sBzK region.

\section{Conclusions}
\label{sec:conclusions}

We have presented a method for constructing end-to-end mock galaxy
catalogues by applying a semi-analytical model of galaxy formation 
to the halo merger trees extracted from a cosmological N-body simulation. 
The mocks that we construct are \emph{lightcone} catalogues, in 
which a galaxy is placed according to the epoch at
which it first enters the past lightcone of the observer. Thus our
catalogues incorporate the evolution of galaxy properties
that is predicted over the simulation snapshots. We use interpolation 
to determine the positions of galaxies at epochs intermediate to the 
simulation snapshots, which represents an improvement over previous 
work. We have shown that our adopted interpolation scheme leads to 
accurate predictions for real space galaxy clustering down to scales well
within the one-halo regime.

We can summarise our method for constructing lightcone catalogues as
follows:
\begin{enumerate}
\item Populate the dark matter halos in the snapshot outputs of 
  a cosmological N-body simulation with galaxies using a physical 
  model of galaxy formation, giving populations of galaxies 
  at a range of cosmic
  epochs. Here we use the dark matter halos from the \emph{Millennium
  Simulation} \citep{Springel05a}, which we populate with galaxies,
  whose positions and properties are calculated using the \GALFORM{}
  semi-analytical model. (In this work we adopt the \cite{Bower06}
  version of \GALFORM{}.)
\item Position an observer within the simulation box.  Replicate the
  simulation box to span a cosmological volume that is of sufficient
  size to encompass the galaxy survey that we wish to mimic.
\item For replication of the box, use adjacent pairs of simulation
  snapshots to determine the epoch at which each galaxy enters the
  observer's past lightcone. Use interpolation to determine the
  corresponding position of the galaxy at this epoch. Reject all
  galaxies that enter the observer's lightcone at a position outside
  of the solid-angle of the galaxy survey.
\item Assign each galaxy that enters the lightcone the intrinsic
  properties that the galaxy had at the lowest redshift snapshot prior
  to the galaxy entering the lightcone. Use the position of the galaxy
  to convert luminosities and absolute magnitudes into observed fluxes and
  apparent magnitudes. Reject all galaxies that fall outside of the
  flux limits which define the galaxy survey.
\end{enumerate}

Our approach has a number of attractive features. First, we use 
a physic model of galaxy formation which makes {\it ab initio} predictions. 
This means that we can build mocks for epochs or selections which 
are currently unprobed. Empirical approaches are not able to do this, 
as they depend on the existence of observations. 
Second, our construction method is generic and is not tied to 
a particular choice of N-body simulation or semi-analytic
model. As better N-body simulations or more accurate galaxy formation 
models become available, our method can still be used. 
Third, the semi-analytic model that we have used has a unique multi-wavelength 
capability, which means that we can mimic surveys built using many different 
telescopes such as GAMA. 

As an illustrative application of our method we considered the effectiveness of
the BzK colour selection technique which is designed to isolate
galaxies within the redshift range $1.4<z<2.5$ \citep{Daddi04b}. The
aim of this exercise is to determine how successful this technique is
at isolating galaxies within the target redshift range and whether the
galaxies it selects are representative of the target population or a
biased subsample. 

The \GALFORM{} model is able to match reasonably well the ${\rm
K}$-band number counts of all BzK galaxies, as well as the counts of
sBzK galaxies. However, the model under-predicts the number of bright
pBzK galaxies and over-predicts the number of faint pBzK galaxies. The
latter discrepancy is partially due to the effect of the depth of
${\rm B}$-band photometry, but may also be related to the crude
estimate of the stripping of gas from satellite galaxies that is
carried out in the \cite{Bower06} model. The BzK technique
successfully selects the majority of the galaxy population within
$2\lesssim z<2.5$ (and possibly out as far as $z \sim 3$), though is
less efficient for $1.4<z\lesssim 2.0$. Examination of the
effectiveness of the BzK technique as a function of ${\rm K}$-band
limiting magnitude suggests that the technique recovers $\gtrsim 75$
per cent of the $1.4<z<2.5$ galaxy population for ${\rm K}$-band
limits fainter than ${\rm K_{AB}}\sim 22$. For brighter limits the
completeness decreases substantially as the BzK population becomes
dominated by low redshift interlopers with $z\leqslant 1.4$.  For
magnitude limits ${\rm K_{\AB}}\gtrsim 21.5$, the fraction of
contamination from BzK galaxies outside $1.4<z<2.5$, remains
approximately constant at $\sim 30$ per cent. We have also shown that
a variation in the typical ${\rm B}$-band magnitude across the BzK
plane can lead to the mis-classification of pBzK galaxies as sBzK
galaxies if the ${\rm B}$-band photometry is of insufficient
depth. Finally, we considered the intrinsic properties of BzK
galaxies, including their stellar mass, SFR, metallicity and stellar
mass weighted age. We find that BzK galaxies display distributions of
these various properties that are in good agreement with the
corresponding distributions for all galaxies with ${\rm
K_{\AB}}\gtrsim 20.5$. However, at brighter ${\rm K}$-band limits BzK
galaxies appear to be less massive, more star-forming, less metal-rich
and younger than the overall population. This is likely related to the
under-prediction of the bright pBzK number counts. The presence of
dust increases the scatter in the colours of (faint) sBzK galaxies,
though does not dramatically change the colour distribution of
galaxies within $1.4<z<2.5$.

We conclude that the BzK colour selection does provide a
representative sample of the $1.4<z<2.5$ population, working better
for fainter ${\rm K}$-band flux limits. However, the depth of ${\rm
B}$-band photometry and extinction due to dust may lead to confusion
between the sBzK and pBzK subsets. 

The tool that we have developed in this paper is a valuable resource
to aid in the exploitation of a wide range of surveys, from
traditional optical selection to novel properties, such as the neutral
hydrogen content of galaxies. Lightcone mock catalogues for different
surveys will be made available for download at
\href{http://www.dur.ac.uk/a.i.merson/lightcones.html}{http://www.dur.ac.uk/a.i.merson/lightcones.html}.

\section*{Acknowledgments}

We thank David Murphy and Daniel Farrow for many useful discussions
and suggestions and Emanuele Daddi for provision of the B, z and K
filter response curves. AIM acknowledges the support of a Science and
Technologies Facilities Council (STFC) studentship. PN acknowledges a
Royal Society URF and an ERC StG grant (DEGAS-259586). Calculations
for this paper were performed on the ICC Cosmology Machine, which is
part of the DiRAC Facility jointly funded by STFC, the Large
Facilities Capital Fund of BIS and Durham University. This work was
supported by a rolling grant from the STFC.

\bibliographystyle{mn2e_trunc8}
\bibliography{aimerson}

\end{document}